\documentclass[a4paper,12pt]{article}
\usepackage{graphicx}
\usepackage{amssymb}
\usepackage{amsmath}
\begin{document}

\newtheorem{t1}{Theorem}[section]
\newtheorem{d1}{Definition}[section]
\newtheorem{c1}{Corollary}[section]
\newtheorem{l1}{Lemma}[section]
\newtheorem{r1}{Remark}[section]

\newcommand{\cA}{{\cal A}}
\newcommand{\cB}{{\cal B}}
\newcommand{\cC}{{\cal C}}
\newcommand{\cD}{{\cal D}}
\newcommand{\cE}{{\cal E}}
\newcommand{\cF}{{\cal F}}
\newcommand{\cG}{{\cal G}}
\newcommand{\cH}{{\cal H}}
\newcommand{\cI}{{\cal I}}
\newcommand{\cJ}{{\cal J}}
\newcommand{\cK}{{\cal K}}
\newcommand{\cL}{{\cal L}}
\newcommand{\cM}{{\cal M}}
\newcommand{\cN}{{\cal N}}
\newcommand{\cO}{{\cal O}}
\newcommand{\cP}{{\cal P}}
\newcommand{\cQ}{{\cal Q}}
\newcommand{\cR}{{\cal R}}
\newcommand{\cS}{{\cal S}}
\newcommand{\cT}{{\cal T}}
\newcommand{\cU}{{\cal U}}
\newcommand{\cV}{{\cal V}}
\newcommand{\cX}{{\cal X}}
\newcommand{\cW}{{\cal W}}
\newcommand{\cY}{{\cal Y}}
\newcommand{\cZ}{{\cal Z}}

\def\cl{\centerline}
\def\bd{\begin{description}}
\def\be{\begin{enumerate}}
\def\ben{\begin{equation}}
\def\benn{\begin{equation*}}
\def\een{\end{equation}}
\def\eenn{\end{equation*}}
\def\benr{\begin{eqnarray}}
\def\eenr{\end{eqnarray}}
\def\benrr{\begin{eqnarray*}}
\def\eenrr{\end{eqnarray*}}
\def\ed{\end{description}}
\def\ee{\end{enumerate}} 
\def\al{\alpha}
\def\b{\beta}
\def\bR{\bar\R}
\def\bc{\begin{center}}
\def\ec{\end{center}}
\def\d{\dot}
\def\D{\Delta}
\def\del{\delta}
\def\ep{\epsilon}
\def\g{\gamma}
\def\G{\Gamma}
\def\h{\hat}
\def\iny{\infty}
\def\La{\Longrightarrow}
\def\la{\lambda}
\def\m{\mu}
\def\n{\nu}
\def\noi{\noindent}
\def\Om{\Omega}
\def\om{\omega}
\def\p{\psi}
\def\pr{\prime}
\def\r{\ref}
\def\R{{\bf R}}
\def\ra{\rightarrow}
\def\s{\sum_{i=1}^n}
\def\si{\sigma}
\def\Si{\Sigma}
\def\t{\tau}
\def\th{\theta}
\def\Th{\Theta}

\def\vep{\varepsilon}
\def\vp{\varphi}
\def\pa{\partial}
\def\un{\underline}
\def\ov{\overline}
\def\fr{\frac}
\def\sq{\sqrt}

\def\WW{\begin{stack}{\circle \\ W}\end{stack}}
\def\ww{\begin{stack}{\circle \\ w}\end{stack}}
\def\st{\stackrel}
\def\Ra{\Rightarrow}
\def\R{{\mathbb R}}
\def\bi{\begin{itemize}}
\def\ei{\end{itemize}}
\def\i{\item}
\def\bt{\begin{tabular}}
\def\et{\end{tabular}}
\def\lf{\leftarrow}
\def\nn{\nonumber}
\def\va{\vartheta}
\def\wh{\widehat}
\def\vs{\vspace}
\def\Lam{\Lambda} 
\def\sm{\setminus}
\def\ba{\begin{array}}
\def\ea{\end{array}} 
\def\bd{\begin{description}}
\def\ed{\end{description}}
\def\lan{\langle}
\def\ran{\rangle}

\baselineskip 16truept

\bc
{\bf Combinatorial approach to multipartite quantum systems : basic formulation} \\

\vspace{.2in} 

{\bf Ali Saif M. Hassan\footnote{Electronic address: alisaif@physics.unipune.ernet.in} and  Pramod Joag\footnote{Electronic address: pramod@physics.unipune.ernet.in}}\\
 Department of Physics, University of Pune, Pune, India-411007. 

\ec

 In this paper we give a method to associate a graph with an arbitrary density matrix referred to a standard orthonormal basis in the Hilbert space of a finite dimensional quantum system. We study the related issues like classification of pure and mixed states, Von-Neumann entropy, separability of multipartite quantum states and quantum operations in terms of the graphs associated with quantum states. In order to address the separability and entanglement questions using graphs, we introduce a modified tensor product of weighted graphs, and establish its algebraic properties. In particular, we show that Werner's definition [1] of a separable state can be written in terms of a graphs, for the states in a real or complex Hilbert space. We generalize the separability criterion (degree criterion)  due to S.L. Braunstein, S. Ghosh, T. Mansour, S. Severini, R.C. Wilson [2], to a class of weighted graphs with real  weights. We have given some criteria for the Laplacian associated with a weighted graph to be positive semidefinite.


\section{Introduction} 

Quantum information is a rapidly expanding field of research because of its theoretical advances in fast algorithms, superdence quantum coding, quantum error correction, teleportation, cryptography and so forth [3,4,5]. Most of these applications are based on entanglement in quantum states.  Although entanglement in pure state systems is relatively well understood, its understanding in the so called mixed quantum states [6], which are statistical mixtures of pure quantum states, is at a primitive level.  Recently, Braunstein, Ghosh and Severini [2, 7], have initiated a new approach towards the mixed state entanglement by associating graphs with density matrices and understanding their classification using these graphs. Hildebrand, Mancini and Severini  [8] testified that the degree condition is equivalent to the PPT-criterion. They also considered the concurrence of density matrices of graphs and pointed out that there are examples on four vertices whose concurrence is a rational number. In this paper we generalize the work of these authors and give a method to associate a graph with the density matrix (real or complex), of an arbitrary density operator, and also to associate a graph with the matrix representing hermitian operator (observable) of the quantum system, both with respect to a standard orthonormal basis in Hilbert space.  We define a modified tensor product of graphs and use it to give Werner's definition for the separability of a $m$-partite quantum system, in  $\mathbb{R}^{q_1} \otimes \mathbb{R}^{q_2} \otimes \cdots \otimes \mathbb{R}^{q_m}$, as well as  $\mathbb{C}^{q_1} \otimes \mathbb{C}^{q_2} \otimes \cdots \otimes \mathbb{C}^{q_m}$ in terms of graphs. We also deal with classification of pure and mixed states and related concepts like Von-Neumann entropy in terms of graphs.

The paper is organized as follows.  In Section 2, we define weighted graphs and their generalized Laplacians which correspond to density matrices, and discuss the permutation invariance of this association.  We also deal with pure and mixed states in terms of graphs.  Section 3 deals with Von-Neumann entropy.  Section 4 is concerned with separability issues as mentioned above.  In Section 5, we deal with graph operations which correspond to quantum operations [5, 9, 10].  In Section 6, we present a method to associate a graph with a general hermitian matrix, having complex off-diagonal elements. We define the modified tensor product for complex weighted graphs and express the separability of mixed quantum states in a complex Hilbert space in terms of graphs, using Werner's definition. In section 7, we present some graphical criteria for the associated Laplacian to be positive semidefinite. Finally, we close with  a summary and some general comments. Sections 2 to 5 deal with graphs with real weights , that is, quantum states living in real Hilbert space. Graphs with  complex weights, corresponding to density operators with complex off diagonal elements are treated in section 6. However, a large part of the results obtained for real Hilbert space in sections 2 to 5, go over to the case of complex Hilbert space (see section 8 (ix)).

\section{Density matrix of a weighted graph} 

\subsection{Definitions}

 A graph $G = (V, E)$ is a pair of a nonempty and finite set called vertex set $V(G)$, whose elements are called vertices and a set $E(G)\subseteq V^2(G)$ of unordered pairs of vertices called edges.  A loop is an edge of the form $\{ v_i, v_i\}$ for some vertex $v_i$.  A graph $G$ is on $n$ vertices if $|V(G)| = n.$  We call the graph as defined above a simple graph.  $|E(G)| = m+s$, where  $m$ : number  of edges joining vertices,  $s$ : number  of loops in $G$ [11].  

A {\it weighted graph} $(G, a)$ is a graph together with a {\it weight function} [12]
$$ a : V(G) \times V(G) \ra I\!\!R$$ 
which associates a  real number (weight) $a(\{u, v\})$ to each pair $\{u, v\}$ of vertices.  The function $a$ satisfies the following properties: 

\bd
\i(i) $a(\{u, v\}) \ne 0$ if $\{u, v\} \in E(G, a)$ and $a(\{u, v\}) = 0$ if $\{u, v\} \not\in E(G, a)$.

\i(ii) $a(\{u, v\}) = a(\{v, u\})$ 

\i(iii) $a(v, v) \ne 0$ if $\{v,v\}\in E(G,a)$ and is zero otherwise. 
\ed 

If $e = \{u, v\}$ is an edge in $E(G, a)$, property (ii) allows us to write $a(e)$ or $a_{uv}$ for $a(\{u, v\})$.  A simple graph can be viewed as a weighted graph with all   nonzero weights equal to 1. 

In the case of simple graphs the degree $d_v$ of a vertex $v \in V(G)$ is defined as the number of edges in $E(G)$ incident on $v$.  For a weighted graph we set 

\ben
   d_{(G,a)} (v) =d_v = \sum_{u \in V(G,a)} a_{uv}.  \label{try}
 \een
  
The {\it adjacency matrix} of a weighted graph with $n$ vertices $M(G, a) = [a_{uv}]_{u, v \in V(G, a)}$ is a $n \times n$ matrix whose rows and columns are indexed by vertices in $V(G,a)$ and whose $uv$-th element is $a_{uv}$.  Obviously the adjacency matrix $M(G, a)$ is a real symmetric matrix with diagonal element $vv$ equal to the weight of the  loops on vertex $v$ (i.e $a_{vv})$.

The {\it degree matrix} for the weighted graph $\D(G, a)$ is a $n \times n$ diagonal matrix, whose rows and columns are labeled by vertices in $V(G, a)$ and whose diagonal elements are the degrees of the corresponding vertices. 
$$\D(G, a) = diag [d_v; v \in V(G, a)]. \eqno{(2)}$$ 
The {\it combinatorial Laplacian} of a weighted graph is defined to be 
$$ L(G, a) = \D(G, a) - M(G, a). \eqno{(3)}$$ 

The {\it degree sum} of $(G, a)$ is defined as 
$$d_{(G, a)} = \sum_{v \in V(G, a)} d_v = Tr \D(G, a) . \eqno{(4)}$$ 
The Laplacian defined by (3) has no record of loops in the graph.  Therefore we define the {\it generalized Laplacian} of a graph $(G, a)$, which includes loops, as
$$ Q(G, a) = \D(G, a) - M(G, a) + \D_0(G, a) \eqno{(5)}$$ 
where $\D_0(G, a)$ is a $n \times n$ diagonal matrix with diagonal elements equal to the weights of the loops on the corresponding vertices 
$$ [\D_0(G, a)]_{vv} = a_{vv} . \eqno{(6)}$$ 
We call $\D_0(G, a)$ the {\it loop matrix} of the graph $(G, a)$ . 

For a given weighted graph $(G, a)$, the generalized Laplacian, defined by (5), is not necessarily a positive semidefinite matrix.  When, for a given graph $(G, a)$, the generalized Laplacian $Q(G,a)$ is positive semidefinite, we can define the density matrix of the corresponding graph $(G, a)$ as 
$$ \si(G, a) = \fr{1}{d_{(G, a)}} Q(G, a) = \fr{1}{d_{(G, a)}} [L(G, a) + \D_0(G, a)] \eqno{(7)}$$ 
where $Tr(\si(G, a)) = 1$. 
Note that, this definition of the density matrix of a weighted graph $(G, a)$ reduces to that of the density matrix for a simple graph without loops [7]. 

Whenever we can define the density matrix for a graph $(G, a)$ we say that the graph $(G, a)$ has density matrix. 

For any density matrix $\si$, we can obtain the corresponding graph as follows: 

\bd
\i(i) Determine the number of vertices of the graph from the size $(n \times n)$ of the density matrix.  The number of vertices = $n$.  Label the vertices from 1 to $n$. 

\i(ii) If the $ij$-th element of $\si$ is not zero draw an edge between vertices $v_i$ and $v_j$ with weight $- \si_{ij}$. 

\i(iii) Ensure that $d_{v_i} = \si_{ii}$ by adding loop of appropriate weight to $v_i$ if necessary. 
\ed

{\it Example}(1): For the following three matrices,we find the corresponding graphs.  

\bd 
\i(1) $ \si = \fr{1}{2} \left[ \ba{cc} 1 & 1 \\ 1 & 1 \ea \right] $
in $\mathbb{R}^2$.  
\begin{figure}[!h]
\begin{center}
\includegraphics[width=3cm,height=.5cm]{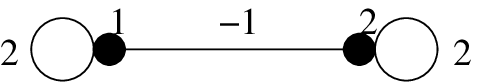}
%\caption{\label{fig:epsart}

Figure 1
\end{center}
\end{figure}

\i(2) $\si = \fr{1}{16} \left[ \ba{rrrr} 9 & -1 & -1 & 1 \\ -1 & 3 & -1 & -1 \\ -1 & -1 & 3 & -1 \\ 1 & -1 & -1 & 1 \ea \right] $ 
in  $\mathbb{R}^2 \otimes \mathbb{R}^2$.
\begin{figure}[!h]
\begin{center}
\includegraphics[width=4cm,height=3cm]{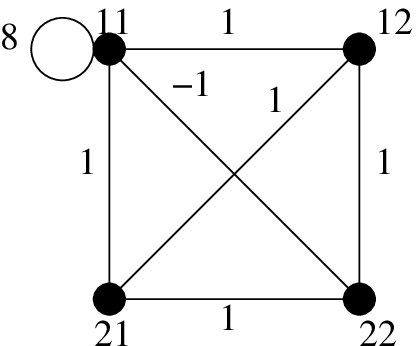}

Figure 2
\end{center}
\end{figure}

\i(3) $\si = \fr{1}{4} \left[ \ba{cccc} 1 & 0 & 0 & 0 \\ 0 & 1 & 0 & 0 \\ 0 & 0 & 1 & 0 \\ 0 & 0 & 0 & 1 \ea \right] $
in  $\mathbb{R}^2 \otimes \mathbb{R}^2$.
 \begin{figure}[!h]
 \begin{center}
\includegraphics[width=4cm,height=4cm]{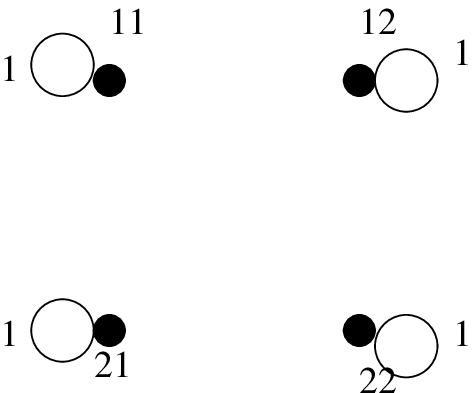}

Figure 3
\end{center}
\end{figure}

\ed 

\subsection{Invariance under isomorphism} 

Two weighted graphs $(G, a)$ and $(G', a')$ are isomorphic if there is a bijective map  [13] 
$$ \phi : V(G, a)  \longmapsto V(G', a')$$
such that 
$$ \{\phi(v_i),  \phi(v_j)\} \in E(G' a') ~ \mbox{iff}~~ \{v_i, v_j\} \in E(G, a), i, j = 1, 2, \cdots, n$$ 
and 
$$ a_{\phi(v_i) \phi(v_j)}' = a_{v_iv_j} ~~ i, j = 1, 2, \cdots, n.$$
We denote isomorphism of $(G, a)$ and $(G', a')$ by $(G, a) \cong (G' a')$. 

Equivalently, two graphs $(G, a)$ and $(G', a')$ are isomorphic if there exists a permutation matrix $P$ such that 
$$ P^TM(G, a) P = M(G', a').$$
Note that,  
$$ P^T \D(G, a)P = \D(G', a'); ~~P^T \D_0(G, a)P = \D_0(G', a')$$
Therefore we have 
$$ P^TQ(G, a) P = Q(G', a'). \eqno{(8)}$$
This means that $Q(G, a)$ and $Q(G', a')$ are similar and have the same eigenvalues.  Therefore, if $Q(G, a)$ is positive semidefinite then so is $Q(G', a')$.  Therefore, if $(G, a)$ has the density matrix so does $(G', a')$.  We have proved

\noi {\bf Theorem 2.1 :} The set of all weighted graphs having density matrix is closed under isomorphism.\hspace{\stretch{1}}$ \blacksquare$

Since isomorphism is an equivalence relation, this set is partitioned by it, mutually isomorphic graphs forming the partition.

\subsection{Correspondence with quantum mechanics} 

Henceforth, we consider only the graphs having density matrix unless stated otherwise.   The basic correspondence with QM is defined by the density matrix of the graph.  For a graph with $n$ vertices the dimension of the Hilbert space of the corresponding quantum system is $n$.  To establish the required correspondence we fix an orthonormal basis in the Hilbert space $\mathbb{R}^{q_1} \otimes \mathbb{R}^{q_2} \otimes \cdots \otimes \mathbb{R}^{q_m}$ of the system, which we call the standard basis and denote it by $\{ | ijk\ell \cdots \ran\}, i, j, k, \ell \cdots = 1, 2, \cdots, n = q_1 q_2 \cdots q_m$, or by $\{ |v_i\ran\}, i = 1, \cdots, n = q_1 q_2 \cdots q_m$.  We label $n$ vertices of the graph $(G, a)$ corresponding to the given density matrix by the $n$ basis vectors in the standard basis.  We say that the graph $(G, a)$ corresponds to the quantum state (density operator) whose matrix in the standard basis is the given density matrix.  Finally, we set up a procedure, by associating appropriate projection operators with edges and loops of $(G, a)$ to reconstruct this quantum state from the graph $(G, a)$.  (See  Theorem 2.7). In view of Theorem 2.1 , if $(G,a)$ has density matrix $\si$ and $(G,a)\cong(G',a')$ with the corresponding permutation matrix $P$,then $(G',a')$ has the density matrix $P^T \si P$. All of this paragraph applies to the complex weighted graph (section 6).

\subsubsection{Pure and mixed states} 

A density matrix $\rho$ is said to be pure if $Tr(\rho^2) = 1$ and mixed otherwise.  Theorem  2.3 gives a necessary and sufficient condition on a graph $(G, a)$ for $\si(G, a)$ to be pure. For a graph $(G, a)$  having $k$ components $(G_1, a_1) , (G_2, a_2), \cdots, (G_k, a_k)$ we write $(G, a) = (G_1, a_1) \uplus (G_2, a_2) \uplus \cdots \uplus (G_k, a_k)$ where $a_i, i = 1, \cdots, k$ are the restrictions of the weight function of the graph $(G, a)$ to the components.  We can order the vertices such that $M(G, a) = \oplus^k_{i=1} M(G_i, a_i)$. When $k = 1$, $(G, a)$ is said to be connected.  From now on we denote by $\la_1(A), \la_2(A), \cdots, \la_k(A)$ the $k$ different eigenvalues of the Hermitian matrix $A$ in the nondecreasing order.  The set of the eigenvalues of $A$ together with their multiplicities is called the spectrum of $A$ [13, 14, 15].

\noi {\bf Lemma 2.2 :} The density matrix of a graph $(G, a)$ without loops has zero eigenvalue with multiplicity  greater than or equal to the number of components of $(G, a)$ with equality applying  when weight function $a=$constant $> 0$,

\noi {\bf Proof :} Let $(G, a)$ be a graph with $n$ vertices and $m$ edges.  Since $Q(G, a)$ is positive semidefinite, for $x \in I\!\!R^n$ we must have [12] 
$$ x^TQ(G,a)x = \sum^m_{k=1} a_{i_kj_k} (x_{i_k} - x_{j_k})^2 + \sum^s_{t=1} a_{i_ti_t} x^2_{i_t} \ge 0.$$ 
For the graph without loops the above inequality becomes 
$$ x^TQ(G, a)x = \sum a_{i_k j_k} (x_{i_k} - x_{j_k})^2 \ge 0. \eqno{(9)}$$
For $x^T = (1~ 1 ~ \cdots 1)$ we can see $x^T Qx= 0$. This means that $x^T = (1~~ 1~~ 1\cdots 1)$ is an unnormalized eigenvector belonging to the eigenvalue 0 [13].  If there are two components $(G_1, a)$ and $(G_2, a)$ of $(G, a)$, with $n_1, m_1$ and $n_2, m_2$ as the number of vertices and edges in $(G_1, a)$ and $(G_2, a)$ respectively, we can decompose the sum in equation (9)  as 
$$ x^TQ(G, a)x = \sum^{m_1}_{k_1=1} a_{i_{k_1}j_{k_1}} (x_{i_{k_1}} - x_{j_{k_1}})^2 + \sum^{m_2}_{k_2=1} a_{i_{k_2}j_{k_2}} (x_{i_{k_2}} - x_{j_{k_2}})^2 . \eqno{(10)}$$
 For $x^T = (1~1~1 \cdots 1)$ both the terms in (10) vanish.  Now consider two vectors $x_1^T = ( 0 ~ 0 ~ \cdots 0  1 ~ 1 \cdots 1)$ with first $n_1$ components zero and last $n_2$ components 1 and $x_2^{T} = (1~~ 1\cdots 1~~0~~0 \cdots 0)$ with first $n_1$ components 1 and last $n_2$ components zero, $(n_1 + n_2 = n)$.  Obviously the RHS of (10) vanishes for both $x_1$ and $x_2$.   This implies $x_1$ and $x_2$ are two orthogonal eigenvectors with eigenvalue zero.  This means multiplicity of zero eigen value is at least 2 (number of components in $(G, a)).$ 

The equality condition for $a_{uv}=$constant$ > 0 ~~ \forall ~~ \{ u, v\} \in E(G, a)$ is proved in [7]. \hspace{\stretch{1}}$ \blacksquare$

\noi {\bf Theorem 2.3 :} The necessary and sufficient condition for the state given by a graph $(G,a )$ to be pure is 
$$ \sum^n_{i=1} d^2_i + 2 \sum^m_{k=1} a^2_{i_kj_k} = d^2_{(G, a)} \eqno{(11)}$$ 
where $d_i$ is the degree of the vertex $v_i$ , $a_{i_kj_k}$ is the weight of the edge $\{ v_{i_k}, v_{j_k}\},(v_{i_k} \ne v_{j_k})$ and $d_{(G, a)}$ is the degree sum $d_{(G, a)} = \sum\limits^n_{i=1} d_i$. 

\noi {\bf Proof :} Equation (11) is just the restatement of the requirement $Tr(\si^2(G, a)) = 1$, which is the necessary and sufficient condition for the state $\si(G,a)$ to be pure.\hspace{\stretch{1}}$ \blacksquare$

\noi {\bf Lemma 2.4 :} The graph $(G,a)$ for a pure state $\si(G,a)$ has the form $(K_\ell, b) \uplus v_{\ell +1} \uplus v_{\ell +2} \uplus \cdots \uplus v_n$ for some $1 \le \ell \le n$.

\noi {\bf Proof :} Since the state is pure, it has the form 
$$ | \psi \ran = \sum^\ell_{k=1} c_{i_k} |v_{i_k}\ran, 1 \le i_k \le n.$$ 
We can permute the basis vectors to transform this sum to $|\psi\ran = \sum\limits^\ell_{i=1} c_i|v_i\ran$.  That is, the $\ell$ basis kets contributing to the sum in the above equation become the vectors $|v_1\ran, |v_2\ran, \cdots, |v_\ell\ran$ under this permutation.  The resulting density matrix $|\psi\ran \lan \psi|$ has a block of first $\ell$ rows and first $\ell$ columns all of whose elements are nonzero, while all the other elements of density matrix are zero.  The graph corresponding to this density matrix is just the required graph. \hspace{\stretch{1}}$ \blacksquare$

{\it Example (2)} :  We now give important cases of pure state graphs in $\mathbb{R}^2$ which we use later. 
%\bd
\begin{figure}[!h]
\includegraphics[width=3cm,height=0.7cm]{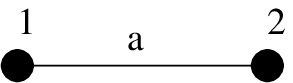}

Figure 4
\end{figure}

(i)$\si(K_2, a) = \fr{1}{2a_{12}} \left[ \ba{cc} a_{12} & - a_{12} \\ -   a_{12} & a_{12} \ea \right] = \fr{1}{2} \left[ \ba{cc} 1 & -1 \\ -1 & 1 \ea \right]   = P[ \fr{1}{\sq 2} (|v_1\ran - |v_2\ran]$, the corresponding graph is as shown in Figure 4.
 \begin{figure}[!h]
\includegraphics[width=3cm,height=1cm]{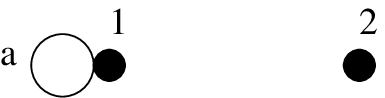}

Figure 5
\end{figure}

(ii) $ \si(K_1, a) = \fr{1}{a} \left[ \ba{cc} a & 0 \\ 0 & 0 \ea \right] = \left[\ba{cc} 1 & 0 \\ 0 & 0 \ea \right] = P[|v_1\ran]$, the corresponding graph is as shown in Figure 5.
\begin{figure}[!h]
\includegraphics[width=3cm,height=1cm]{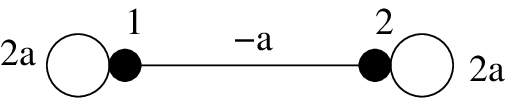}

Figure 6
\end{figure}

(iii)$ a_{12} > 0,  \si(K_2, -a) = \fr{1}{2a_{12}} \left[ \ba{cc} a_{12} & a_{12} \\ a_{12} & a_{12} \ea \right]  = \fr{1}{2} \left[ \ba{cc} 1 & 1 \\ 1 & 1 \ea \right] = P[ \fr{1}{\sq 2} (|v_1\ran + |v_2\ran)], a > 0$. The corresponding graph is as shown in Figure 6.

\begin{figure}[!h]
\includegraphics[width=3cm,height=3cm]{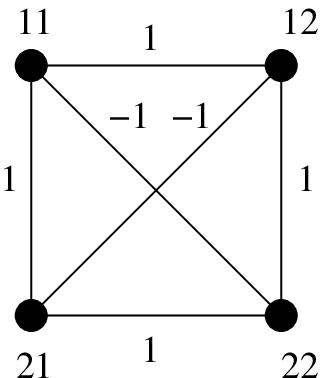}

Figure 7
\end{figure}
(iv) $$ \si(G, a) = \fr{1}{4} \left[ \ba{cccc} 1 & -1 & -1 & 1 \\ -1 &  1 &  1 & -1 \\ -1 & 1& 1 & -1\\ 1 & -1 & -1 & 1 \ea \right]= P[ (|-\ran  |-\ran)]$$, where $ |-\ran=\fr{1}{\sq 2}(|1\ran - |2\ran)$, the corresponding graph is as shown in Figure 7.
 \begin{figure}[!h]
\includegraphics[width=4cm,height=4cm]{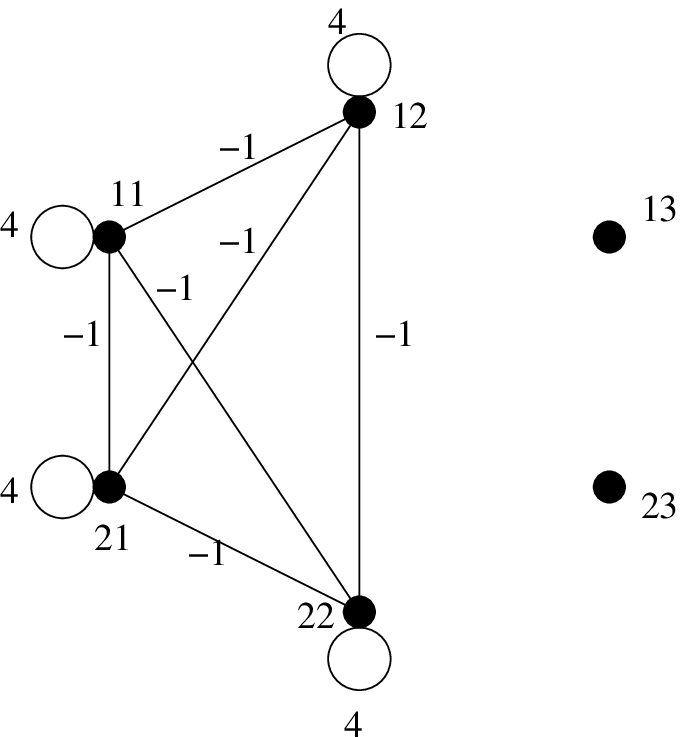}

Figure 8
\end{figure}

(v) $$ \si(G, a) = \fr{1}{4} \left[ \ba{cccccc} 1 & 1 & 0 & 1 & 1 & 0 \\ 1 &  1 & 0 &  1 & 1 & 0 \\ 0 & 0 & 0 & 0 & 0 &0 \\1 & 1& 0 &  1 & 1 & 0 \\ 1 & 1 & 0 & 1 & 1 & 0\\ 0 & 0 & 0 & 0 & 0 & 0 \ea \right]= P[ (|+\ran  |+\ran)],$$ where $ |+\ran=\fr{1}{\sq 2}(|1\ran + |2\ran)$
in $\mathbb{R}^2 \otimes \mathbb{R}^3$, the corresponding graph is as shown in Figure 8.

It may be seen that in each of the cases in example (2), same density matrix on the standard basis corresponds to infinite family of graphs as the  nonzero weight on  each edge or loop is multiplied by a constant.  But this is a false alarm because any weight $a \ne 1$ only changes the length of the corresponding state vector in the Hilbert space (i.e. state becomes unnormalized) which does not have any physical significance.  Another example pertaining to this situation is the random mixture (see Lemma (3.1 )). 
$$ \si(G, a) = \fr{1}{an} \left[ \ba{cccc} a \\ & a & & 0 \\ & & \ddots \\ 0 & & & a \ea \right] = \fr{1}{n} \left[ \ba{cccc} 1 & & & 0 \\ & 1 & & \\ & & 1 \\ & & \ddots \\ 0 & & & 1 \ea \right] = \fr{1}{n} I_n.$$ 
However, this does not lead to any contradiction because of the uniqueness of the random mixture [6]. 

All the  density matrices in (i), (ii), (iii), (iv), (v)   above represent pure states. 

\noi{\bf Remark 2.5 :} Any graph with weight function $a=$ constant $>0$ has the same density matrix for all $a>0$. This infinite family of graphs corresponds to the same quantum state (density operator).

\noi {\bf Definition 2.6 :} A graph $(H, b)$ is said to be a factor of graph $(G, a)$ if $V(H, b) = V(G, a)$ and there exists a graph $(H', b')$ such that $V(H', b') = V(G, a)$ and $M(G, a) = M(H, b) + M(H', b')$.  Thus a factor is only a spanning subgraph.  Note that 
$$ a_{v_iv_j} = \left\{ \ba{lll} b_{v_iv_j} & \mbox{if} & \{v_i,v_j\} \in E(H, b) \\ b'_{v_iv_j} & \mbox{if} & \{v_i,v_j\} \in E(H', b') \ea \right. $$ 

Now let  $(G, a)$ be a graph on $n$ vertices $v_1, \cdots, v_n$ having $m$ edges\\ $\{v_{i_1}, v_{j_1}\}, \cdots, \{v_{i_m}, v_{j_m}\}$ and $s$ loops $\{v_{i_1}, v_{i_1}\} \cdots \{v_{i_s}, v_{i_s}\}$ where $1 \le i_1j_1, \cdots, i_m j_m \le n, 1 \le i_1 i_2 \cdots i_s \le n$.

Let $(H_{i_kj_k}, a_{i_kj_k})$ be the factor of $(G, a)$ such that 
$$ [M(H_{i_kj_k}, a_{i_kj_k})]_{u,w} = \left\{ \ba{l} a_{i_kj_k} ~~ \mbox{if}~~ u = i_k~~ \mbox{and}~~ w = j_k ~~\mbox{or}~~ u = j_k, w = i_k \\ 0  ~~ \mbox{otherwise} \ea \right. \eqno{(12)}$$ 

Let $(H_{i_t,i_t}, a_{i_t i_t})$ be a factor of $(G, a)$ such that 
$$ [M(H_{i_ti_t}, a_{i_t i_t})]_{uw} = \left\{ \ba{l} a_{i_t i_t}~~ \mbox{when}~~ u = i_t = w \\ 0 ~~ \mbox{otherwise} \ea \right. \eqno{(13)}$$ 

\noi {\bf Theorem 2.7 :} The density matrix of a graph $(G, a)$ as defined above with factors given by equation (12) and (13) can be decomposed as 
$$ \si(G, a) = \fr{1}{d_{(G, a)}} \sum^m_{k=1} 2a_{i_kj_k} \si(H_{i_kj_k}, a_{i_kj_k}) + \fr{1}{d_{(G, a)}} \sum^s_{t=1} a_{i_ti_t} \si(H_{i_ti_t}, a_{i_ti_t}) \eqno{(14)}$$ 
or
$$ \si(G, a) = \fr{1}{d_{(G, a)}} \sum^m_{k=1} 2a_{i_kj_k} P[\fr{1}{\sq 2}(|v_{i_k}\ran - |v_{j_k}\ran)] + \fr{1}{d_{(G, a)}}  \sum^s_{t=1} a_{i_ti_t} P[|v_{i_t}\ran]\eqno{(15)}$$

\noi {\bf Proof :} From equation (12), (13) and Theorem 2.3 and Lemma 2.4, the density matrix 
$$\si (H_{i_kj_k}, a_{i_kj_k}) = \fr{1}{2a_{i_kj_k}} [ \D(H_{i_kj_k}, a_{i_kj_k}) - M(H_{i_kj_k},a_{i_kj_k})]$$
is a pure state.  Also, 
$$ \si (H_{i_ti_t}, a_{i_ti_t}) = \fr{1}{a_{i_ti_t}} [ \D_0 (H_{i_t, i_t}, a_{i_ti_t})]$$
is a pure state.  Now 
$$ \D(G, a) = \sum^m_{k=1} \D(H_{i_kj_k}, a_{i_kj_k}) + \sum^s_{t=1} \D_0(H_{i_ti_t}, a_{i_ti_t})$$
$$M(G, a) = \sum^m_{k=1} M(H_{i_kj_k}, a_{i_kj_k}) + \sum^s_{t=1} \D_0(H_{i_ti_t}, a_{i_ti_t}).$$
Therefore 
%\rr
$$\si(G, a)  =  \fr{1}{d_{(G, a)}} \left[ \sum^m_{k=1} \D(H_{i_kj_k}, a_{i_kj_k}) - \sum^m_{k=1} M(H_{i_kj_k}, a_{i_kj_k})\right]\\
 + \fr{1}{d_{(G, a)}} \left[ \sum^s_{t=1} \D_0 (H_{i_ti_t}, a_{i_ti_t})\right]$$
$$=  \fr{1}{d_{(G, a)}} \sum^m_{k=1} [\D(H_{i_kj_k}, a_{i_kj_k}) - M(H_{i_kj_k}, a_{i_kj_k})] \\
 + \fr{1}{d_{(G, a)}} \sum^s_{t=1} \D_0(H_{i_ti_t}, a_{i_ti_t})$$
$$ =  \fr{1}{d_{(G, a)}} \sum_k 2a_{i_kj_k} \si(H_{i_kj_k}, a_{i_kj_k}) \\
 + \fr{1}{d_{(G, a)}} \sum_t a_{i_ti_t} \si(H_{i_ti_t}, a_{i_ti_t}) ~~~~~~~~~~\mbox{(14)}$$

 In terms of the standard basis, the $uw$-th element of matrices $\si(H_{i_kj_k}, a_{i_kj_k})$ and $\si(H_{i_ti_t}, a_{i_ti_t})$ are given by $\lan v_u | \si(H_{i_kj_k} , ,a_{i_kj_k}) | v_w \ran$ and $\lan v_u | \si (H_{i_ti_t} a_{i_ti_t} | v_w\ran$ respectively.  In this basis 
$$ \si(H_{i_kj_k}, a_{i_kj_k}) = P[ \fr{1}{\sq 2} ( | v_{i_k} \ran - | v_{j_k} \ran )]$$ 
$$ \si(H_{i_ti_t}, a_{i_ti_t}) = P[| v_{i_t} \ran ] .$$ 

Therefore equation (14) becomes 
%\benr
$$\si(G, a)  =  \fr{1}{d_{(G, a)}} \sum^m_{k=1} 2a_{i_kj_k} P[\fr{1}{\sq 2} (| v_{i_k}\ran - | v_{j_k} \ran) + \fr{1}{d_{(G, a)}} \sum ^s_{t=1} a_{i_ti_t} P[ | v_{i_t} \ran]~~~~~~~~~~\mbox{(15)}$$
%\eenr
$\hspace{\stretch{1}} \blacksquare$

\noi {\bf Remark 2.8 :} If all weights $a_{i_kj_k} > 0$ then equations (14), (15) give $\si(G, a)$ as a mixture of pure states.  However, in the next subsection we show that any graph $(G, a)$ having density matrix can be decomposed into graphs (spanning subgraphs) corresponding to pure states.

\subsubsection{Convex combination of density matrices}

Consider two graphs $(G_1, a_1) $ and $(G_2, a_2)$ each on the same $n$ vertices, having $\si(G_1, a_1)$ and $\si(G_2, a_2)$ as their density matrices respectively.  We give an algorithm to construct the graph $(G, a)$ whose density matrix is 
$$ \si(G, a) = \la \si(G_1, a_1) + (1 - \la) \si(G_2, a_2)$$
$0 \le \la \le 1, \la = \al/\b, \al, \b > 0 $ are real. 

We use the symbol $\sqcup$ to denote the union of the edge sets of two graphs $(G_1, a_1) $ and $(G_2, a_2)$ on the same set of vertices to give $(G,a).$ If $\{v_i,v_j\} \in E(G_1,a_1)$ and $\{v_i,v_j\} \in E(G_2,a_2)$ then $a(\{v_i,v_j\}) = a_1(\{v_i,v_j\}) + a_2(\{v_i,v_j\})$. We write $(G,a)= (G_1, a_1) \sqcup(G_2, a_2).$
 If $E(G_1,a_1)$ and $E(G_2,a_2)$ are disjoint sets, then we call the resulting graph $(G,a)$ the disjoint edge union of $(G_1,a_1)$ and $(G_2,a_2)$, we write $(G,a)= (G_1, a_1) \dotplus(G_2, a_2).$ 

The algorithm is as follows : 

\noi {\bf Algorithm 2.9 :} 

\bd
\i(i) Put $\la = \al/\b$ so that $(1 - \la) = \fr{\b - \al}{\b}$, where $\al>0$ ,$\b>0$ are real.

\i(ii) Write $\si(G, a) = \fr{1}{\b} (\al \si(G_1,a_1) + (\b - \al) \si(G_2, a_2))$.

\i(iii) Modify the weight functions of the two graphs $(G_1, a_1)$ and $(G_2, a_2)$ to get $a_1' = \al a_1$ and $a_2' = (\b - \al)a_2$.

\i(iv) The graph $(G, a)$ corresponding to $\si$ in step (ii) is 
$$ (G, a) = (G_1, a_1') \sqcup (G_2, a_2') \eqno{(16)}$$
such that 
$$ a_{v_iv_j} = (a_1')_{v_iv_j} + (a_2')_{v_iv_j} \eqno{(16a)}$$
$$ a_{v_iv_i} = (a_1')_{v_iv_i} + (a_2')_{v_iv_i} \eqno{(16b)}$$
where we take $(a_{1,2}')_{v_iv_j} = 0 = (a_{1,2}')_{v_iv_i}$  if $\{v_i,v_j\},\{v_i,v_i\} \not\in E(G_1, a_1)$ or $E(G_2, a_2)$

\ed 
$\hspace{\stretch{1}}$ $\blacksquare$

We can apply this algorithm to any convex combination of more than two density matrices $\si(G, a)= \sum\limits^k_{i=1} p_i \si(G_i, a_i),\; \sum\limits_i p_i = 1$, by writing $p_i = \al_i/\b, \al_i,\b >0$ and real$, i = 1, \cdots, k$.

 {\it Example(3)} : consider the density matrices

\bd
\i(i) $\si(G_1, a_1) = | + + \ran \lan + + | = \fr{1}{4} \left[ \ba{cccc} 1 & 1 & 1 & 1\\ 1 & 1 & 1 & 1 \\ 1 & 1 & 1 &1 \\ 1 & 1 & 1 & 1 \ea \right]$\\
whose graph is shown in Figure 9\\
\begin{figure}[!h]
\includegraphics[width=4cm,height=3cm]{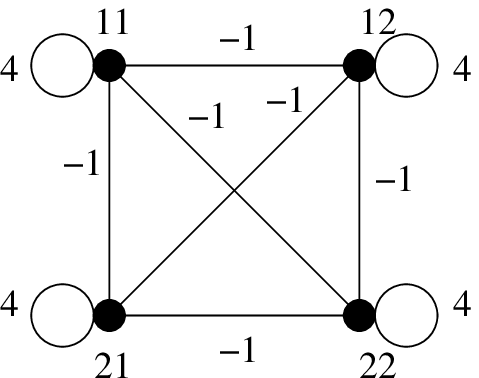}

Figure 9
\end{figure}
\benrr
\si(G_2, a_2) & = & \fr{1}{2} | 11 \ran \lan 11 | + \fr{1}{2} | \psi^+ \ran \lan \psi^+ | \\
& = & \fr{1}{4} \left[ \ba{cccc} 2 & 0 & 0 & 0 \\ 0 & 1 & 1 & 0 \\ 0 & 1 & 1 & 0 \\ 0 & 0 & 0 & 0 \ea \right].
\eenrr
whose graph is shown in Figure 10
\begin{figure}[!h]
\includegraphics[width=4cm,height=3cm]{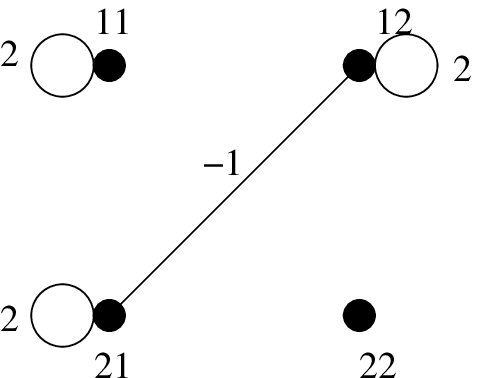}

Figure 10
\end{figure}

The graph corresponding to 
\benrr
\si(G, a) & = & \fr{1}{3}  \si(G_1, a_1) + \fr{2}{3} \si(G_2, a_2) \\
& = & \fr{1}{12} \left[ \ba{cccc} 5 & 1 & 1 & 1 \\ 1 & 3 & 3 & 1 \\ 1 & 3 & 3 & 1 \\ 1 & 1 & 1 &1 \ea \right].
\eenrr
is given in Figure 11
\begin{figure}[!h]
\includegraphics[width=4cm,height=3cm]{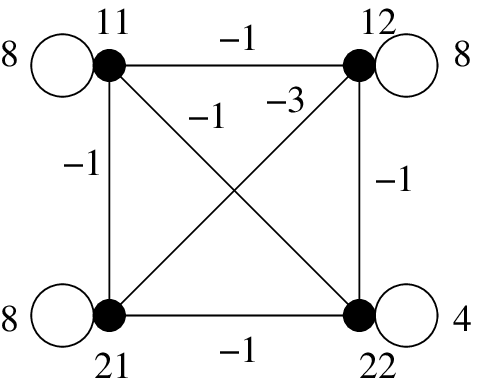}

Figure 11
\end{figure}

\ed 

\noi {\bf Lemma 2.10 :} Let $(G_1, a_1)$, $(G_2, a_2)$ and  $(G, a)$ satisfy 

$$(G, a) = (G_1, a_1) \sqcup (G_2, a_2) $$  or,
$$(G,a)= (G_1, a_1) \dotplus(G_2, a_2).$$ 

Then 
$$ Q(G, a) = Q(G_1, a_1) + Q(G_2, a_2) $$ 
and 
$$ \si(G, a) = \fr{d_{(G_1,a_1)}}{d_{(G, a)}}   \si(G_1, a_1) + \fr{d_{(G_2,a_2)}}{d_{(G, a)}}   \si(G_2, a_2).$$

\noi {\bf Proof :} For two factors of $(G, a), (G_1, a_1)$ and $(G_2, a_2)$ we have 
\benrr
M(G, a) & = & M(G_1, a_1) + M(G_2, a_2) \\
\D(G, a) & = & \D(G_1, a_1) +  \D(G_2, a_2)\\
\D_0(G, a) & = & \D_0(G_1, a_1) + \D_0(G_2, a_2) \\
L(G, a) & = & \D(G, a) - M(G, a)\\ 
Q(G, a) & = & L(G, a) + \D_0(G, a)
\eenrr
Substitute  $M(G, a), \D(G, a), \D_0(G, a)$ and $L(G, a)$ in $Q(G, a)$ as above to get 
$$Q(G, a) = Q(G_1, a_1) + Q(G_2, a_2)$$
and also
$$ \si(G, a) = \fr{d_{(G_1, a_1)}}{d_{(G, a)}} \si(G_1, a_1) + \fr{d_{(G_2, a_2)}}{d_{(G, a)}} \si(G_2, a_2).$$\hspace{\stretch{1}}$ \blacksquare$

\noi {\bf Remark 2.11 :} Obviously, the operation $\sqcup$ is associative. We can apply  Lemma 2.10 for more than two graphs, 
 $$(G, a) = \sqcup_i (G_i, a_i) \Ra Q(G, a) = \sum_i Q(G_i, a_i)$$ 
and 
$$ \si(G, a) = \fr{1}{d_{(G, a)}} \sum_i d(G_i, a_i) \si(G_i, a_i). $$ 

\noi {\bf Theorem 2.12 :} Every graph $(G, a)$ having density matrix $\si(G, a)$ can be decomposed as $(G, a) = \sqcup_i (G_i, a_i)$ where $\si(G_i, a_i)$ is a pure state. 

\noi {\bf Proof :} Every density matrix can be written as the convex combination of pure states $\si(G, a) = \sum\limits^k_{i=1} p_i | \psi_i \ran \lan \psi_i|$.

By applying Algorithm (2.9), Lemma 2.10 and Remark 2.11, we get the result.\hspace{\stretch{1}}$ \blacksquare$

\subsubsection{Tracing out a part}
Consider a bipartite system with dimension $pq$. Let $\si(G,a)$ be a state of the system with graph $(G,a)$ having $pq$ vertices labeled by $(ij), i=1,\cdots, p $ and $j=1, \cdots, q$. If we trace out the second part with dimension $q$, we get the state of the first part which is $p \times p$ reduced density matrix of $\si(G,a)$. The corresponding graph $(G',a')$ has $p$ vertices indexed by $(i)$ and its weight function $a'$ is given by $$ a'_{ij}=\sum_{k=1}^q a_{ik,jk} , i\neq j$$ and $$a'_{ii}=\sum_{k=1}^q d_{ik}-\sum_{l\in V(G',a')} a'_{il}, l\neq i.$$ Where $d_{ik}$ is the degree of vertex $(ik)$ in original graph.

{\it Example (4)} : Consider a graph $(G,a)$ as shown in Figure 12a in $\mathbb{R}^2 \otimes \mathbb{R}^2$. The corresponding density matrix is $$\si^{AB}(G, a) =  \fr{1}{16} \left[ \ba{cccc} 9 & -1 & -1 & 1\\ -1 & 3 & -1 & -1 \\ -1 & -1 & 3 & -1 \\ 1 & -1 & -1 & 1 \ea \right]$$\\
after tracing out the second particle the graph $(G',a')$ on two vertices becomes as in Figure 12b with corresponding density matrix $$\si^A(G', a') = \fr{1}{16} \left[ \ba{cc} 12 & -2\\ -2 & 4 \ea \right]=\fr{1}{8} \left[\ba{cc} 6 & -1\\ -1 & 2\ea \right]$$\\
which is the same as the reduced density matrix $\si^A$ of $\si^{AB}$.

\begin{figure}[!h]
\includegraphics[width=4cm,height=4cm]{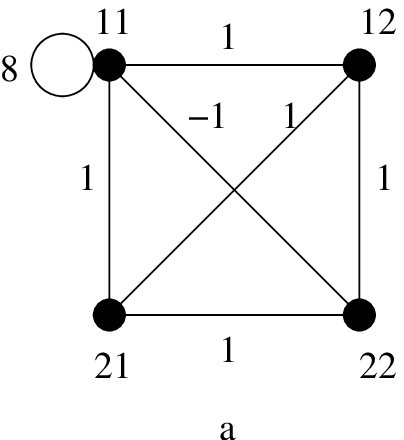}

Figure 12a
\end{figure}

 \begin{figure}[!h]
\includegraphics[width=4cm,height=1cm]{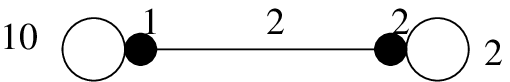}

Figure 12b
\end{figure}

\section{Von Neumann entropy} 

The Von Neumann entropy of $n \times n$ density matrix $\si$ is 
$$ S(\si) = - \sum^n_{i=1} \la_i(\si) \log_2 \la_i(\si) $$
It is conventional to define $0 \log 0 = 0$.  The Von Neumann entropy is a measure of mixedness of the density matrix.  For a pure state $\si, S(\si)= 0$.

\subsection{Maximum and minimum}

  Let 
$$ (G, a) = \uplus^n_{i=1} (K^i_1, a_i) \eqno{(17)}$$ 
where $(K^i_1, a_i)$ is the graph on $i$-th vertex with a loop having weight $a_i > 0$. 

\noi {\bf Lemma 3.1 :} Let $(G, a)$ be given by (17) with the additional constraint that $a_i = c = \fr{1}{n}, i = 1, 2, \cdots, n$.  The density matrix of the graph $(G, a)$ is the random mixture of pure states with $\si(G, a) =  \fr{1}{n} I_n$.

\noi {\bf Proof :} For the graph $(G, a)$, the first term in (14) vanishes. Then 
$$ \si(G, a) = \fr{1}{d_{(G,a)}} \sum^n_{t=1} \D_0(H_{i_ti_t}, a)$$ 
where $\D_0(H_{i_ti_t}, a)$ is the $n \times n$ matrix with all elements zero except $(i_t, i_t)th$ element which is equal to $a$. This means 
$$ \si(G, a) = \fr{a}{d_{(G,a)}} \left[ \ba{cccc} 1 & & & 0 \\ & 1 & & \\ & & \ddots \\ 0 & & & 1 \ea \right] = \fr{1}{n} I_n,$$
because $d_{(G,a)}=na$.\hspace{\stretch{1}}$ \blacksquare$

\noi {\bf Theorem 3.2 :} Let $(G, a)$ be a graph on $n$ vertices. Then 
\bd
\i(i) $\max_{(G, a)} S(\si(G, a)) = \log_2 n$ 

\i(ii) $\min_{(G,a)} s(\si(G, a)) = 0$, and this value is attained if $\si(G, a)$ is pure. 
\ed

\noi {\bf Proof :} (i) By Lemma 3.1 $\si(G, a)$ defined in the Lemma has eigenvalue $1/n$ with multiplicity $n$.  The corresponding Von Neumann entropy is $\log_2 n$.  Since $(G, a)$ is on $n$ vertices, the support of $\si(G, a)$ has dimension $\le n$.  Any matrix having dimension of support $\le n$ cannot have Von Neumann entropy $> \log_2 n$.

(ii) For pure state $S(\si(G,a))=0$ and $S(\si(G,a))\nless 0$ .\hspace{\stretch{1}}$ \blacksquare$

\section{Separability} 

In this section we primarily deal with the graphs representing a bipartite quantum system with Hilbert space $\mathbb{R}^p \otimes \mathbb{R}^q$ of dimension $pq$. Obviously, the corresponding graph has $n = pq$ vertices.  We label the vertices using standard (product) basis  $\{| v_i \ran = | u_{s+1} \ran \otimes |w_t \ran \}, 0 \le s \le p-1, 1 \le t \le q, i = sq + t$. 

\subsection{Tensor product of weighted graphs} 

The tensor product of two graphs $(G, a)$ and $(H, b)$ denoted $(G, a) \otimes (H, b)$ is defined as follows.

The vertex set of $(G, a) \otimes (H, b)$ is $V(G, a) \times V(H, b)$. Two vertices $(u_1, v_1)$ and $(u_2, v_2)$  are adjacent if $\{u_1, u_2\} \in E(G, a)$ and $\{v_1, v_2\} \in E(H, b)$.  The weight of the edge $\{ (u_1, v_1), (u_2, v_2)\}$ given by $a_{\{u_1, u_2\}} b_{\{v_1,v_2\}}$ and is denoted by $c(\{ (u_1, v_1), (u_2, v_2)\})$.  Note that either $u_1$ and $u_2$ or $v_1$ and $v_2$ or both can be identical, to include loops.

The adjacency, degree and the loops matrices of $(G, a) \otimes (H, b)$ are given by 
$$M((G, a) \otimes (H, b)) = M(G, a) \otimes M(H, b) \eqno{(18a)}$$
$$ \D((G, a) \otimes (H, b)) = \D(G, a) \otimes \D(H, b) \eqno{(18b)}$$ 
$$ \D_0((G, a) \otimes (H, b)) = \D_0(G, a) \otimes \D_0(H, b) \eqno{(18c)}$$ 
Note that 
$$ L((G, a) \otimes (H, b)) \ne L(G, a) \otimes L(H, b) $$ 
$$Q((G, a) \otimes (H, b)) \ne Q(G, a) \otimes Q(H, b).$$
In fact, in general, the tensor product of two graphs having density matrix may not have density matrix. 

For two simple graphs $G$ and $H$ we know that [16, 7]
$$d_{G \otimes H} = d_G \cdot d_H.$$
This result is also satisfied by the tensor product of the weighted graphs.
$$ d_{(G, a) \otimes (H, b)} = d_{(G, a)} \cdot d_{(H, b)} . \eqno{(19)}$$ 

\subsection{Modified tensor product}

We modify the tensor product of graphs in order to preserve positivity of the generalized Laplacian of the resulting graph. 

Given a graph $(G, a)$ we define $(G^\phi, a)$ by 
$$ V(G^\phi, a) = V(G, a)$$ 
$$E[(G^\phi, a)] = E(G, a)\setminus \{ \{v_i,v_i\} :  \{v_i, v_i\} \in E(G, a)\}$$ 
That is $(G^\phi, a)$ is obtained from $(G, a)$ by removing all loops.

Given a graph $(G, a)$ we define $(\widetilde{G}, a)$ by 
$$ V(\widetilde{G}, a) = V(G, a)$$ 
$$E(\widetilde{ G}, a) = E(G, a) \setminus \{ \{v_i, v_j\} : i \ne j, \{v_i, v_j\} \in E(G, a)\}.$$
That is, $(\widetilde{ G}, a)$ is obtained by removing all edges connecting neighbors and keeping loops. 

Note that in both $(G^\phi, a)$ and $(\widetilde{ G}, a)$, weight function $a$ remains the same, only its support is restricted.

Given a graph $(G, a)$ we define $(-G, a) = (G, - a)$.  Given a graph $(G, a)$ we define $(G^\#, a')$
$$V(G^\#,a')= V(G,a)$$
$$(G^\#,a')=\uplus_i^n(K_i,a'_i)$$
where $K_i$ is the graph consisting of $i$th vertex with a loop and $a'_i$ is the weight of the loop on the $i$th vertex. If $a'_i=0$ then there is no loop on the $i$th vertex. $a'_i, i=1,2,\cdots n$ are given by $$a'_i=\sum_{v_k \in V(G,a)}a(\{v_i,v_k\})\eqno{(20a)}$$
Note that the term $v_k=v_i$ is also included in the sum.

We now define the graph operators on the set of graphs 
$$ \left. \ba{ll}  \mbox{(i)} & \eta : (G, a) \ra (- G, a) = (G ,-a) \\ \mbox{(ii)} & \cL : (G, a) \ra (G^\phi, a) \\ \mbox{(iii)} & \cN : (G, a) \ra (G^\#, a') \\ \mbox{(iv)} & \Om : (G, a) \ra (\widetilde{ G}, a) \ea \right\} \eqno{(20b)}$$ 

Some properties of the graph operators defined in (20b) are, 
$$ \ba{ll} \mbox{(i)} & M(\eta(G, a)) = - M(G, a) \\ & \D(\eta(G, a)) = - \D(G, a) \\ & \D_0(\eta(G, a)) = - \D_0(G, a) \ea \eqno{(21)}$$ 
$$ d_{\eta(G, a)} = - d_{(G, a)} $$ 
$$ \ba{ll} \mbox{(ii)} & M(\cL(G, a)) = M(G, a) - \D_0(G, a)  \\ & \D(\cL(G, a)) = \D(G, a) - \D_0(G, a)  \ea \eqno{(22)}$$ 
$$ \D_0(\cL(G, a)) = [0] $$ 
$$ d_{\cL(G, a)} = Tr(\D(G, a)) - Tr(\D_0(G, a)) = d_{(G^\phi, a)}$$
$$ \ba{ll} \mbox{(iii)} & M(\cN(G, a)) = \D(G, a) \\ & \D(\cN(G, a)) = \D(G, a) \\ & \D_0(\cN(G, a)) = \D(G, a)  \\ & d_{\cN(G, a)} = Tr(\D(G, a)) = d_{(G, a)} \ea \eqno{(23)}$$ 
$$ \ba{ll} \mbox{(iv)} &  M(\Om(G, a)) = \D_0(G, a) \\& \D(\Om(G, a)) =\D_0(G, a) \\& \D_0(\Om(G, a)) = \D_0(G, a) \\ &  d_{\Om(G, a)} = Tr(\D_0(G, a)) \ea \eqno{(24)}$$ 

{\it Example (5)} :
Given a graph $(G,a)$ as shown in Figure 13a, if we act by $\eta$,   
 $\cL$ ,$\cN$ and $\Om$ on $(G,a)$, we get the graphs $\eta(G,a)$, $\cL(G,a)$, $\cN(G,a)$
 and $\Om(G,a)$
 as shown in Figures 13b, 13c, 13d and 13e respectively.

\begin{figure}
\includegraphics[width=12cm,height=5cm]{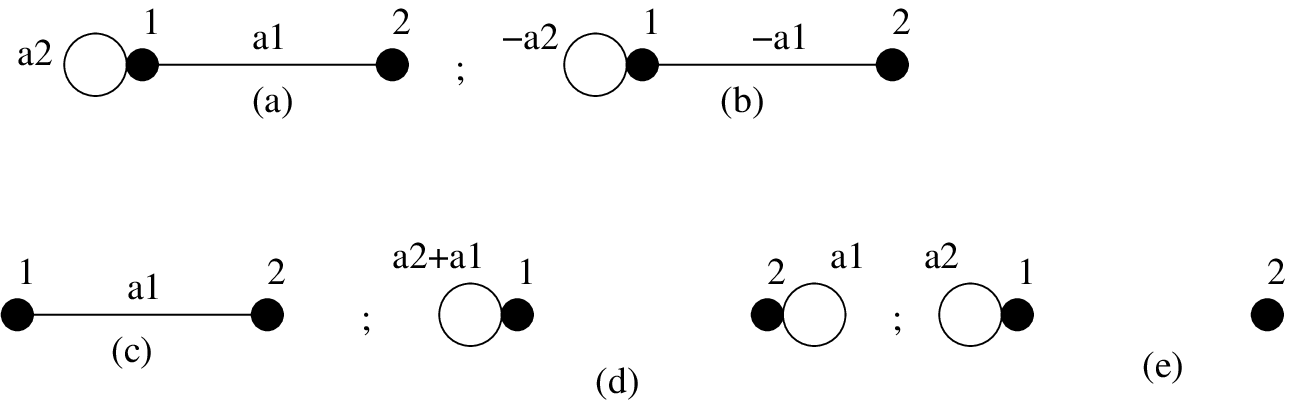}

Figure 13
\end{figure}

\noi {\bf Definition 4.1 :} Let $(G, a)$ and $(H, b)$ be two graphs with $p$ and $q (> p)$ vertices respectively.  Then their modified tensor product is defined by 
\benrr
(G, a) \boxdot (H, b) & = &\{ \cL(G, a) \otimes \cL \eta (H, b)\} \dotplus \{\cL(G, a) \otimes \cN(H, b)\}\\
& & \dotplus \{ \cN(G, a) \otimes \cL(H, b)\} \dotplus \{ \Om(G, a) \otimes \Om(H, b)\}~~~~~~~~~\mbox{(25)} 
\eenrr 
$$ V\{ (G, a) \boxdot (H, b)\} = V(G, a) \times V(H, b) $$
whose cardinality is $pq$. 

$E\{ (G, a) \boxdot (H, b) \} $ = Disjoint union of the edge set of each term in (25).

\noi {\bf Lemma 4.2 :} (i)  $\D((G, a) \boxdot (H, b)) = \D(G, a) \otimes \D(H, b)$.

(ii) $\D_0((G, a) \boxdot (H, b)) = \D_0(G, a) \otimes \D_0(H, b)$.

\noi {\bf Proof :} Consider the degree matrix of the modified tensor product we have
\benrr
 \D((G, a) \boxdot (H, b))& =& \D(\cL(G, a) \otimes \cL\eta(H, b)) + \D(\cL(G,a) \otimes \cN(H, b))\\ 
& & + \D(\cN(G, a) \otimes \cL(H, b)) + \D(\Om(G, a) \otimes \Om(H, b)) \\
\eenrr
This follows from Lemma 2.10. Using  equation (18b) and equations (21) to (24)  to the terms on the RHS of the above equation  we get 

$$\D((G, a) \boxdot (H,b)) = \D(G, a) \otimes \D(H, b).$$ 

Equations (ii) can be proved similarly.\hfill $\blacksquare$

\noi {\bf Corollary 4.3 :} $d_{(G,a) \boxdot (H,b)}(v_1,v_2)\; =\;d_{(G,a)}(v_1) \cdot d_{(H,b)}(v_2)$

\noi {\bf Proof :} This follows directly from equation (i) in Lemma 4.2.\hfill $\blacksquare$

\noi {\bf Remark 4.4 :} From corollary  we get $d_{(G,a)\boxdot(H,b)}\; =\;d_{(G,a)} \cdot d_{(H,b)}$

\noi {\bf Theorem 4.5 :} Consider a bipartite system in $\mathbb{R}^p \otimes \mathbb{R}^q$ in the state $\si$. Then $\si = \si_1 \otimes \si_2$ if and only if $\si$ is the density matrix of the graph $(G, a) \boxdot (H, b)$, where $(G, a)$ and $(H, b)$ are the graphs having density matrices $\si_1$ and $\si_2$ respectively.

\noi {\bf Proof :} \noi {\bf If part :} Given $(G, a), (H, b)$ we want to prove 
$$\si((G, a) \boxdot (H, b)) = \si_1(G, a) \otimes \si_2(H, b).$$

 From the definition of the modified tensor product we can write 
$$ \si((G,a) \boxdot (H, b)) = \fr{1}{d_{(G, a) \boxdot (H, b)}} \{Q[\cL(G, a) \otimes \cL \eta(H, b)$$
$$ \dotplus \cL(G, a) \otimes \cN(H, b)\dotplus  \cN(G, a) \otimes \cL(H, b) \dotplus \Om(G, a) \otimes \Om(H, b)]\}$$
Using Lemma 2.10, Remark 2.11  and Remark 4.4 we get 
\benrr
\si((G, a) \boxdot (H, b)) & = & \fr{1}{d_{(G, a)} \cdot d_{(H, b)}} [ Q(\cL(G, a) \otimes \cL \eta(H, b)) \\
& & + Q(\cL(G, a) \otimes \cN(H, b)) + Q(\cN(G, a)\\
& &  \otimes \cL(H, b)) + Q(\Om(G, a) \otimes \Om(H, b))] ~~~~~~~~~~~~~~~~~~~~~~ \mbox{(26)}
\eenrr
We can calculate every term in (26) using (21) - (24) and substitute in (26) to get 
$$ \si((G, a) \boxdot (H, b)) = \si(G, a) \otimes \si(H, b) .$$ 

\noi {\bf Only if part :} Given $\si = \si_1 \otimes \si_2$, consider the graphs $(G, a)$ and $(H, b)$ for $\si_1$ and $\si_2$ respectively.  Then the graph of $\si$ has the generalized Laplacian 
\benrr
& & [L(G, a) + \D_0(G, a)] \otimes [L(H, b) + \D_0(H, b)] \\
& & = L(G, a) \otimes L(H, b) + L(G, a) \otimes \D_0(G, a) + \D_0(G, a) \otimes L(H, b) + \D_0(G, a) \otimes \D_0(H, b)
\eenrr
 Now it is straightforward to check that the graphs corresponding to each term are given by the corresponding terms in the definition of $(G, a) \boxdot (H, b)$. \hspace{\stretch{1}}$ \blacksquare$

 \noi {\bf Remark 4.6 :} Note that the proof of Theorem 4.5 does not depend in any way on the positivity or the hermiticity of the associated generalized Laplacians. Therefore we have $$Q((G,a) \boxdot(H,b))\;=\;Q(G,a)\otimes Q(H,b)$$ for any two graphs $(G,a)$ and $(H,b)$

\noi {\bf Corollary 4.7 :} The modified tensor product is associative and distributive with respect to the disjoint edge union $\dotplus$.

\noi {\bf Proof :} Let $(G_1,a_1), (G_2,a_2)$ and $(G_3,a_3)$ be any graphs. Using Theorem 4.5 and Remark 4.6 , we can write 
$$Q(((G_1,a_1) \boxdot(G_2,a_2))\boxdot(G_3,a_3))\;$$
$$=\;Q((G_1,a_1) \boxdot(G_2,a_2)) \otimes Q(G_3,a_3) $$
$$=\;(Q(G_1,a_1) \otimes Q(G_2,a_2))\otimes Q(G_3,a_3) $$
 $$=\;Q(G_1,a_1) \otimes (Q(G_2,a_2)\otimes Q(G_3,a_3)) $$
$$=\;Q(G_1,a_1) \otimes Q((G_2,a_2)\boxdot (G_3,a_3) )$$
$$=\;Q((G_1,a_1)\boxdot ((G_2,a_2)\boxdot (G_3,a_3) )$$
 Therefore , 
$$((G_1,a_1) \boxdot(G_2,a_2))\boxdot(G_3,a_3)\;=\;(G_1,a_1)\boxdot ((G_2,a_2)\boxdot(G_3,a_3) )$$

Similarly, using Lemma 2.10 and distributive property of the matrix tensor product we get

$$Q((G_1,a_1) \boxdot((G_2,a_2))\dotplus (G_3,a_3)))\;$$
$$=\;Q((G_1,a_1) \boxdot(G_2,a_2)) \dotplus ((G_1,a_1)  \boxdot (G_3,a_3)) $$
 Which gives 
$$(G_1,a_1) \boxdot ((G_2,a_2)) \dotplus (G_3,a_3))\;=\; (G_1,a_1) \boxdot (G_2,a_2) \dotplus (G_1,a_1)  \boxdot (G_3,a_3) $$

\noi {\bf Definition 4.8 :} The cartesian product of two weighted graphs $(G,a)$ and $(H,b)$  is denoted $(G,a) \square (H,b)$ with weight function $c$ defined as follows.
$$V(G, a) \times V(H, b).$$ 
$E((G,a ) \square (H,b))=\{\{(u,v),(x,y)\} |\; u=x$ and $\{v,y\} \in E(H,b),\; v \ne y,\; c(\{(u,v),(u,y)\}) = d_u \cdot b(\{v,y\}) $ or  $v = y$ and $\{u,x\} \in E(G,a), \; u \ne x,\; c(\{(u,v),(x,v)\})= d_v \cdot a(\{u,x\}).$

Where $d_u$ and $d_v$ are the degrees of the vertices $u \in E(G,a)$ and $v \in E(H,b)$ respectively. It is straightforward to check that 

$$(G,a) \square (H,b)\;=\; \cL(G, a) \otimes \cN(H, b) \dotplus \cN(G, a) \otimes \cL(H, b)$$

Which can be taken to be the definition of the cartesian product of graphs in terms of the operators $\cL$ and $ \cN$ . We also note that 

\benrr
(G, a) \boxdot (H, b) & = &\{ \cL(G, a) \otimes \cL \eta (H, b)\} \dotplus \{(G,a) \square (H,b)\} \dotplus \{ \Om(G, a) \otimes \Om(H, b)\} 
\eenrr 
 
Note that the isolated vertices in $(G,a)$ or $(H,b)$ do not contribute to $(G,a) \square (H,b)$ as their degree is zero.

{\it Example(6)} : Consider $(G,a), (H,b)$ where $V(G,a)=\{1,2\}, E(G,a)=\{ \{1,2\}\} $ and $V(H,b)=\{1,2,3,4\}, E(H,b)=\{\{1,2\}, \{2,3\}, \{3,4\}\}$ with weight functions $a=b=1$,  as shown in Figure $14a,14b$. The modified tensor product of these graphs is given by Figures 15a, 15b, 15c, 15d, for each term in (25) and the resulting graph is as shown in Figure 15e. The corresponding density matrix of the graph $(G,a)\boxdot(H,b)$ is

$$\si((G, a)\boxdot (H,b)) \; = \; \fr{1}{12} \left[ \ba{cccccccc} 1 & -1 & 0 & 0 & -1 & 1 &   0 & 0 \\-1 & 2 & -1 & 0 & 1 & -2 & 1 & 0 \\ 0 & -1 & 2 & -1 & 0 & 1 & -2 & 1 \\ 0 & 0 & -1 & 1 & 0 & 0 & 1 & -1 \\ -1 & 1 & 0 & 0 & 1 & -1 & 0 & 0 \\ 1 & -2 & 1 & 0 & -1 & 2 & -1 & 0 \\ 0 & 1 & -2 & 1 & 0 & -1 & 2 & -1 \\ 0 & 0 & 1 & -1 & 0 & 0 & -1 & 1 \ea \right].$$

 which is the same as  $\si(G,a)\otimes \si (H,b)$ .

\begin{figure}[!h]
\includegraphics[width=6cm,height=3cm]{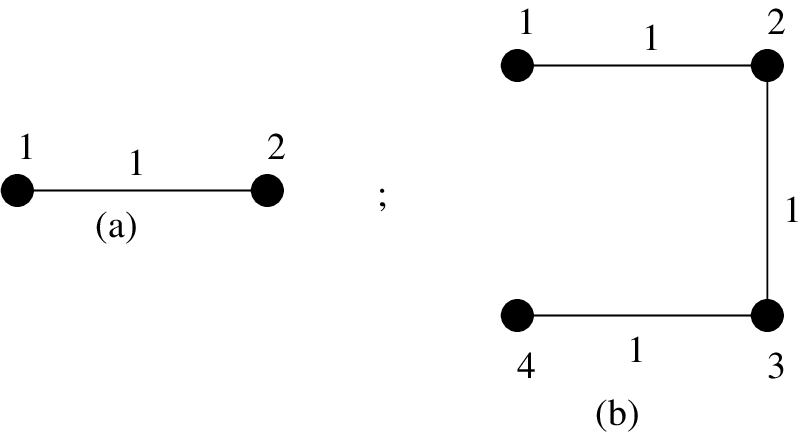}

Figure 14
\end{figure}

\begin{figure}[!h]
\includegraphics[width=8cm,height=6cm]{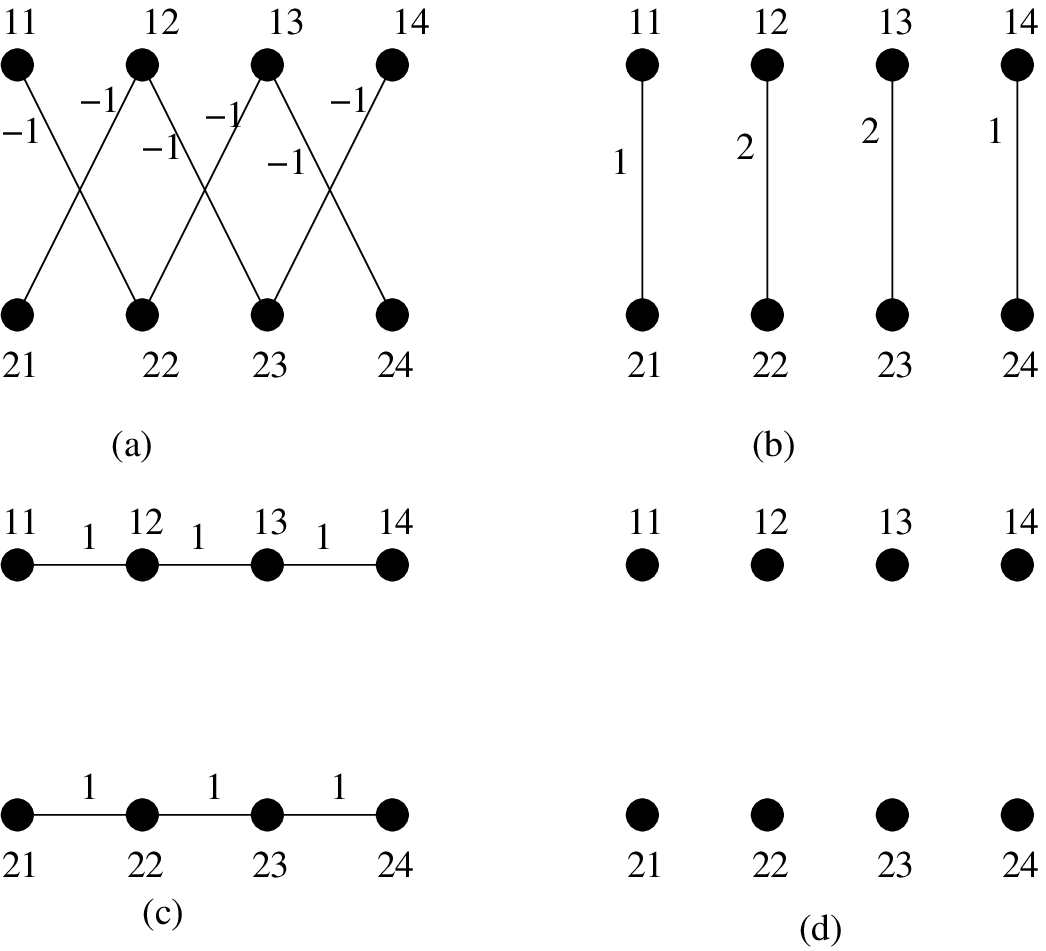}

\includegraphics[width=4cm,height=3cm]{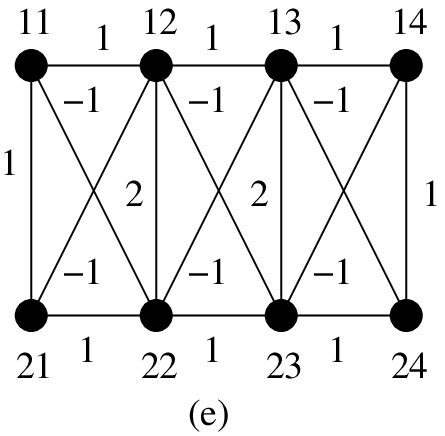}

Figure 15

\end{figure} 

\noi {\bf Corollary 4.9 :} The density matrix of the modified tensor product of two graphs is separable.

\noi {\bf Proof :} From Theorem 4.5 we see that $\si((G, a) \boxdot (H, b))$ is actually a product state. \hspace{\stretch{1}}$ \blacksquare$

\noi {\bf Corollary 4.10 :} $\si = \si_1 \otimes \si_2 \otimes \cdots \otimes \si_k$ for a $k$-partite system if and only if the graph of $\si$ is the modified tensor product of the graphs of $\si_1, \cdots, \si_k$. 

\noi {\bf Proof :} Apply Theorem 4.2 successively to $(\si_1 \otimes \si_2), ((\si_1 \otimes \si_2) \otimes \si_3)\cdots $ and then use the associativity of the modified tensor product corollary 4.7. \hspace{\stretch{1}}$ \blacksquare$

\noi {\bf Corollary 4.11 :} A state $\si$ of a $k$-partite system is separable if and only if the graph (G,a)  for  $\si$ has the form 
$$ (G,a)\;=\; \sqcup_i \boxdot^k_{j=1} (G^j_i, a^j_i). $$

\noi {\bf Proof :} Let $\si$ be separable i.e. 
$$ \si = \sum_i w_i \si_i^{(1)} \otimes \si_i^{(2)} \otimes \cdots \otimes \si_i^{(k)}, ~~ \sum_i w_i = 1.$$ 

By Algorithm 2.9 and Corollary 4.10 the graph of $\si$ has the form 
$$(G, a) = \sqcup_i \boxdot^k_{j=1} (G^j_i, a^j_i).$$
Now let the graph of a $k$-partite state be 
$$(G, a) = \sqcup_i \boxdot^k_{j=1} (G^j_i, a^j_i).$$ 
Then by Lemma 2.10 , Remark 2.11 and the above corollary to Theorem 4.5 
$$ \si(G, a) = \sum_i w_i \si_1^{(1)} \otimes \si_1^{(2)} \otimes \cdots \otimes \si_i^{(k)}.$$ \hspace{\stretch{1}}$ \blacksquare$

  Corolary 4.11 says that Werner's definition [1] of a separable state in  $\mathbb{R}^{q_1} \otimes \mathbb{R}^{q_2} \otimes \mathbb{R}^{q_3} \otimes \cdots \otimes \mathbb{R}^{q_k}$ system,  can be expressed using corresponding graphs.

\noi {\bf Lemma 4.12 :} For any $n = pq$ the density matrix $\si(K_n, a)$ is separable in $\mathbb{R}^p \otimes \mathbb{R}^q$ if the weight function is constant $> 0$.

\noi {\bf Proof :} The proof is same as that given for corollary 4.3 in [7], for simple graph.\hspace{\stretch{1}}$ \blacksquare$

{\it Example (7)} : Consider the graph $(K_4, a)$.  The vertices of $(K_4, a)$ are denoted by 1, 2, 3, 4, where weight function is constant, say , $a = 3 > 0$ and has loops in vertices 1, 2.  We associate to these vertices the orthonormal basis $\{ |1\ran = |1\ran |1\ran, |2\ran = |1\ran |2\ran, |3\ran = |2\ran |1\ran, |4\ran = |2\ran |2\ran\}$.  In terms of this basis $\si(K_4, a)$ can be written as 
$$ \si(K_4, a) = \fr{1}{42} \left[ \ba{rrrr} 12 & -3 & -3 & -3 \\ -3 & 12 & -3 & -3 \\ -3 & -3 & 9 & -3 \\ -3 & -3 & -3 & 9 \ea \right] = \fr{1}{14} \left[ \ba{rrrr} 4 & -1 & -1 & -1 \\ -1 & 4 & -1 & -1 \\ -1 & -1 & 3 & -1 \\ -1 & -1 & -1 & 3 \ea \right] $$
and from equation (15) we can write $\si(K_4, a)$ as 
\benrr
\si(K_4, a) & = & \fr{1}{42} [6P[|1\ran \fr{1}{\sq 2} (|1\ran - |2\ran)] + 6P[ \fr{1}{\sq 2} (|1\ran - |2\ran)|1\ran]\\
& & + 6P[ \fr{1}{\sq 2} (|1\ran - |2\ran) |2\ran] + 6P[|2\ran \fr{1}{\sq 2} (|1\ran - |2\ran)] + 6P[\fr{1}{\sq 2} (|11\ran - |22\ran)] \\
& & 6P[ \fr{1}{\sq 2} (|12\ran - |21\ran)] + 3P[|11\ran] + 3P[|12\ran]\}
\eenrr
\benrr
\si(K_4, a) & = & \fr{1}{7} P[|1\ran \fr{1}{\sq 2} (|1\ran - |2\ran)] + \fr{1}{7} P[ \fr{1}{\sq 2}(|1\ran - |2\ran) |1\ran] \\
& & + \fr{1}{7} P[ \fr{1}{\sq 2} ( |1\ran - |2\ran)|2\ran] + \fr{1}{7} P[|2\ran \fr{1}{\sq 2} (|1\ran - |2\ran)] \\
& & + \fr{2}{7} \{ \fr{1}{2} P[ \fr{1}{\sq 2} (|11\ran - |22\ran)] + \fr{1}{2} P[ \fr{1}{\sq 2} (|12\ran - |21\ran)]\} \\
& & \fr{1}{14} P[|11\ran] + \fr{1}{14} P[|12\ran].
\eenrr
Each of the first four terms in the above expression is a projector on a product state, and also the last two terms are  projectors, while the fifth and sixth terms give rise to the separable density matrix $\fr{1}{2} p[\mid - \ran \mid + \ran] + \fr{1}{2} p[| + \ran | - \ran]$, where $| \pm\ran \stackrel{def}{=} \fr{1}{\sq 2} (|1\ran \pm |2\ran)$ [7].  Thus $\si(K_4, a)$ , $a$ constant, is separable in $\mathbb{R}^2 \otimes \mathbb{R}^2$. 

Note that there exists a graph which is complete with a  real weight function, which is entangled as the  following graph shows in 

\begin{figure}[!h]
\includegraphics[width=2cm,height=2cm]{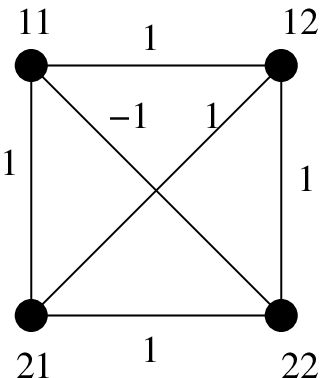}

Figure 16
\end{figure}

\noi {\bf Remark 4.13 :} Separability of $\si(K_n, a)$ with constant weight function $> 0$ does not depend upon the labeling of $V(K_n, a)$ provided every vertex has  a loop or there are no loops.  Given a graph, an isomorphism from $(G, a) \longmapsto (G, a)$ is called automorphism.  Under composition of maps, the set of automorphisms of $(G, a)$ form a group denoted $Aut(G, a)$.  If $\si(K_n, a)$ is separable, and if the $Aut(K_n, a) = S_n ,\; (G, a) \cong (K_n, a)$ is also separable. Note that $Aut(K_n, a) = S_n$ provided all weights are equal and either every vertex has a loop or there are no loops.

\noi {\bf Lemma 4.14 :} The complete graph $(K_n, a)$ on $n \geq 2$ vertices corresponding to a separable state with weight function $ constant > 0$ is not a modified tensor product of two graphs. 

\noi {\bf Proof :} It is clear that, if $n$ is prime then $(K_n, a)$ is not a  modified tensor product of graphs.  We then assume that $n$ is not a prime.  Suppose that there exist graphs $(G, b)$ and $(H, c)$ respectively on $p$ and $s$ vertices such that $(K_{ps}, a) = (G, b) \boxdot (H, c)$ where $ b$ and $c$ are constants .  From the definition of the modified tensor product 
$$ a(\{ (u_1,v_1), (u_2, v_2)\}) = b(\{ u_1, u_2\}) \cdot c(\{ v_1, v_2\}),$$
the degree sum is
$$ d_{(G, b)} = \sum_{u \in V(G, b) } d_u = \sum_{u \in V(G, b)} \sum_{w \in V(G, b)} b_{wv} = 2b|E(G, b)|.$$
We know that 
$ d_{(G, b)} \le b(p(p-1))$ and also $d_{(H, c)} = 2c|E(H, b)| \le cs(s-1)$ and 
$$ d_{(G, b)}\cdot d_{(H, c)} \le bcps(p-1) (s-1) = bcps(ps - p - s + 1) .\eqno{(27)}$$ 
Now observe that $V((G, b) \boxdot (H, c)) = ps$ and, 
$$ d_{(G, b) \boxdot (H, c)} = aps(ps -1) \eqno{(28)}$$ 
because $(G, b) \boxdot (H, c) = (K_{ps}, a)$. 

We know that 
$$ d_{(G, b) \boxdot (H, c)} = d_{(G, b)}\cdot d_{(H, c)}. \eqno{(29)}$$ 
 Substituting from (27) and (28) we see that (29) is satisfied only when $ p = 1 = s $ , i.e. $n = ps = 1$. \hspace{\stretch{1}}$ \blacksquare$

Lemma 4.12, Lemma 4.14 and Theorem 4.5 together imply that a complete graph $(K_n,a)$ on $n \ge2$  vertices with $a=$ constant $>0$ is a separable state but  not a product state.

{\bf Definition 4.15 :} Consider a graph $(G,a)$, without loops, pertaining to a bipartite system of dimension $pq$ . The partial transpose of $(G,a)$, denoted $(G^{\Gamma_B},a')$ , is a graph defined as $V(G^{\Gamma_B},a')=V(G,a)$ , $\{il,kj\}\in E(G^{\Gamma_B},a') \Longleftrightarrow \{ij,kl\}\in E(G,a)$ and $a'(\{il,kj\}=a(\{ij,kl\}.$

{\bf Lemma 4.16 :} Consider a bipartite separable state $\si(G,a)$ with the associated graph $(G,a)$ without loops. Then $\Delta(G,a)=\Delta(G^{\Gamma_B},a'),$  where $(G^{\Gamma_B},a')$ is the partial transpose of $(G,a)$.
 
{\bf Proof :} Let $Q(G,a)$ be the Laplacian of a graph $(G,a)$ with  real  weights without  loops, on $n$ vertices. Let $D$ be any $n\times n$ real diagonal matrix in the standard orthonormal basis $\{|v_i\ran\};i=1,2,\dots,n$, such that $D\neq 0$ and $Tr(D)=0$. This means that there is at least one negative entry in the diagonal of $D$. Denote this element by $D_{ii}=b_i$. Let $|\psi_0 \ran= \sum_j|v_j\ran$ and $|\phi\ran=\sum_j \chi_j|v_j \ran$  where, 
\begin{displaymath}
\chi_j = 
\left\{ \begin{array}{ll}
 0 & \textrm{if $j\neq i$}\\
k \in R & \textrm{if $j=i$}
\end{array} \right.
\end{displaymath}
Let $|\chi \ran=|\psi\ran+|\phi\ran=\sum_{i=1}^n(1+\chi_j)|v_j\ran$. Then 
\benrr
\lan\chi|Q(G,a)+D|\chi\ran& = & \lan\psi_0|Q(G,a)|\psi_0\ran+
\lan\psi_0|Q(G,a)|\phi\ran+\lan\phi|Q(G,a)|\psi_0\ran+\\
& &              \lan\phi|Q(G,a)|\phi\ran+\lan\psi_0|D|\psi_0\ran+\lan\psi_0|D|\phi\ran+\lan\phi|D|\psi_0\ran+\lan\phi|D|\phi\ran
\eenrr
 Since $|\psi_0\ran$ is (unnormalized) vector having all components equal unity, from equation (9) it follows that $\lan\psi_0|Q(G,a)|\psi_0\ran=0$. Also $\lan\psi_0|D|\psi_0\ran=Tr(D)=0$. 
  We have
 $$\lan\phi|Q(G,a)|\phi\ran=k^2(Q(G,a))_{ii}=k^2d_i$$
 $$\lan\psi_0|Q(G,a)|\phi\ran=\lan\phi|Q(G,a)|\psi_0\ran = 0.$$
 
 Finally, the remaining terms in the above equation are given by 
 $$\lan\phi|D|\phi\ran=b_ik^2$$
 $$\lan\psi_0|D|\phi\ran = b_ik=\lan\phi|D|\psi_0\ran.$$
 Thus
 $$\lan\chi|Q(G,a)+D|\chi\ran = k^2(b_i+d_i)+ 2kb_i$$
  So we can then always choose a positive $k$ , such that
 $$\lan\chi|Q(G,a)+D|\chi\ran < 0.$$
 It then follows $Q(G,a)+D \ngeq 0.$

  This expression is identical with that obtained in [2].
  For any graph $G$ on $n=pq$ vertices $$v_1=u_1w_1, v_2=u_1w_2, \dots, v_{pq}=u_pw_q,$$ consider  the degree condition $\Delta(G) = \Delta(G^{\Gamma_B}).$ Now $$(L(G))^{\Gamma_B}=(\Delta(G) - \Delta(G^{\Gamma_B}))+ L(G^{\Gamma_B}).$$
  Let $$D=\Delta(G) - \Delta(G^{\Gamma_B}).$$
  Then $D$ is an $n\times n$ real diagonal matrix with respect to the orthonormal basis $$|v_i\ran = |u_1\ran\otimes |w_1\ran, \dots, |v_{pq}\ran = |u_p\ran\otimes |w_q\ran.$$
  Also $$Tr(D)=Tr(\Delta(G))-Tr(\Delta(G^{\Gamma_B}))= 0.$$
  We have two possible cases : $D\neq 0$ or $D = 0$. If $D\neq 0$, that is the degree condition is not satisfied $(i.e. \Delta(G)\neq \Delta(G^{\Gamma_B}))$ we have seen that $L(G)+ D\ngeq 0$. As a consequence, $L(G^{\Gamma_B})+D\ngeq 0$ and then $(L(G))^{\Gamma_B}\ngeq 0.$ Hence $\rho(G) $ is entangled.\hspace{\stretch{1}}$ \blacksquare$

 {\bf Lemma 4.17 :} A graph $(G,a)$  for a bipartite state corresponds to a separable state if $\{ij,kl\}$ $(i\neq k,j\neq l) \in E(G,a)\Longrightarrow \{il,kj\}\in E(G,a)$ and $a_{ij,kl}=a_{il,kj}$.
 
 {\bf Proof :} Suppose $a_{ij,kl}=a_{il,kj}=a,i\neq k,j\neq l.$ The contribution of the corresponding two edges is $$a \{ P[\fr{1}{\sq2}(|ij\ran-|kl\ran)]+P[\fr{1}{\sq2}(|il\ran-|kj\ran)]\}$$ which is a separable state. Thus all such pairs contribute separable states.  Any other edge $\{ij,kl\}$ with $i=k$ or $ j=l$ has the contribution $a_{ij,kl}P[|i\ran \otimes (\fr{1}{\sq2}(|j\ran-|l\ran))]$ which is separable.  Loops contribute the product states $P[|ii\ran]$.
\hspace{\stretch{1}}$ \blacksquare$

The reverse implication is not true in general. The counter-example is the graph (Figure 12a) in example (4) which is separable.

\section{Graph Operators} 

A graph operation is a map that takes a graph to another graph [17].  We deal with four cases namely deleting and adding an edge and deleting and adding a vertex. 

Deleting an edge $(\{ v_i, v_j\}, a_{v_iv_j})]$ from a graph $(G, a)$ results in a graph\\ $(G, a) -  ( \{ v_i, v_j\}, a_{v_iv_j}) \stackrel{def}{=} (V(G, a), E(G, a) \setminus \{ v_i, v_j\})$ with $a_{v_iv_j} = 0$.  Note the possibility $v_i = v_j$ corresponding to the edge being a loop.  Addition of an edge $(\{ u_i, v_j\}, a_{ij})$ maps $(G, a)$ to the graph $(G, a) + (\{ v_i, v_j\}, a_{ij}) \stackrel{def}{=} [V(G, a), E(G, a) \cup \{ v_i, v_j\}]$ with $a_{v_iv_j} = a_{ij}$.  Deletion of a vertex $v_i$ maps $(G,a)$ to $(G, a) - \{ v_i\} \stackrel{def}{=} [V(G, a) \setminus \{ v_i\}, E(G, a) \setminus E_i]$ where $E_i$ is the set of all edges incident to $v_i$ (including the loop on $v_i)$ with the weight function zero for the edges in $E_i$.  Adding a vertex $v_i$ to $(G, a)$ maps $(G, a)$ to $(G, a) + \{ v_i\} \stackrel{def}{=} (V(G, a) \cup \{ v_i\}, E(G, a))$.

A very important point is that, in general, the set of graphs having density matrix is not closed under these operations.  Addition of an edge with positive weight and deletion of an edge with negative weight preserves the positivity of the generalized Laplacian resulting in the graph having density matrix.  However, addition (deletion) of an edge with negative (positive) weight may lead to a graph which does not have density matrix.  In the next section, we give a method for addition and deletion of an edge which preserves the positivity of the generalized Laplacian.  Deletion and addition of vertices always preserves the positivity of the generalized Laplacian.

Let $\cB(\cH^n)$ be the space of all bounded linear operators on $\cH^n$.  A linear map $\Lambda : \cB(\cH^n) \ra \cB(\cH^m)$ is said to be hermiticity preserving if for every hermitian operator $O \in \cB(\cH^n), \Lambda(O)$ is an hermitian operator in $\cB(\cH^m)$.  A hermiticity preserving map $\Lambda : \cB(\cH^n) \ra \cB(\cH^m)$ is said to be positive if for any positive operator $O \in \cB(\cH^n), \Lambda(O)$ is a positive operator in $\cB(\cH^m)$.  A positive map $\Lambda : \cB(\cH^n) \ra \cB(H^m)$ is said to be completely positive if for each positive integer $k, (\Lambda \otimes I_{k^2}) : \cB(\cH^n \otimes \cH^k) \ra \cB(H^m \otimes \cH^k)$ is again a positive map.  A completely positive map $\Lambda : \cB(\cH^n) \ra \cB(\cH^m)$ is said to be trace preserving if $Tr(\Lambda(O)) = Tr(O)$, for all $O \in \cB(\cH^n)$.  A quantum operation is a trace preserving completely positive map (for short, TPCP) [9, 5].  In standard Quantum Mechanics, any physical transformation of a quantum mechanical system is described by a quantum operation [6].  We are going to use the following result: 

\noi {\bf (Kraus representation  Theorem)} [10] : Given a quantum operation $\Lambda : \cB(\cH^n) \ra \cB(\cH^m)$, there exist $m \times n$ matrices $A_i$, such that $\Lambda(\rho) = \sum\limits_i A_i \rho A_i^\dagger$, where $\rho$ is any density matrix acting on $\cH^n$ and $\sum\limits_i A^\dagger_i A_i = I_m$ (The converse is true).  The matrices $A_i$'s are called Kraus operators. 

A projective measurement $\cM = \{ P_i; i = 1, 2, \cdots, n\}$, on a quantum mechanical system $S$ whose state is $\rho$, consists of pairwise orthogonal projectors $P_i : \cH_s \ra \cH_s$, such that $\sum\limits^n_{i=1} P_i = I_{dim(\cH_s)}$.  The $i$-th outcome of the measurement occurs with probability $Tr(P_i \rho)$ and the post-measurement state of $S$ is $\fr{P_i\rho P_i}{tr(P_i \rho)}.$ Whenever the $i$-th outcome of the measurement occurs, we say that $P_i$ clicks. Last two paragraphs apply to complex Hilbert space and so also to real Hilbert space.

\subsection{Deletion and addition of an edge for a weighted graph with all weights $> 0$ } 

Here we describe how to delete or add an edge by means of TPCP.  Our method of deleting an edge from a weighted graph with all positive weights is a simple generalization of the method in [7].  Let $(G, a)$ be a graph on $n$ vertices $v_1, \cdots, v_n$ and $m$ edges $\{v_{i_1} v_{j_1}\} \cdots \{v_{i_m} v_{j_m}\}, i_k \ne j_k, k = 1, \cdots, m$ and $s$ loops $\{ v_{i_1} v_{i_1} \} \cdots \{ v_{i_s}v_{i_s}\}$.  Our purpose is to delete the edge $\{ v_{i_k} v_{j_k}\}, i_k \ne j_k$.  Then we have 
\benrr
 \si(G,a) = \fr{1}{d_{(G,a)}} \{ \sum^m_{\ell=1} 2a_{i_\ell j_\ell} P[\fr{1}{\sq 2} (| v_{i_\ell} \ran - | v_{j_\ell} \ran)]  + \sum^s_{t=1} a_{i_ti_t} P[|v_{i_t} \ran ]\} 
\eenrr
and 
\benrr
\si((G, a) - \{ v_{i_k} v_{j_k}\})  =  \fr{1}{d_{(G,a)} - 2a_{i_kj_k}} 
 \left\{ \sum^m_{\ba{c} \ell =1 \\ \ell \ne k \ea} 2a_{i_\ell j_{\ell}} P [ \fr{1}{\sq 2} (|v_{i_\ell} \ran - | v_{j_\ell} \ran)] + \sum^s_{t=1} a_{i_ti_t} P[|v_{i_t}\ran ]\right\}.
\eenrr

A measurement in the basis $\cM = \{ \fr{1}{\sq 2} (| v_{i_k} \ran \pm | v_{j_k} \ran), |v_i\ran : i \ne i_k, j_k$ and $i = 1, 2, \cdots, n\}$ is performed on the system prepared in the state $\si(G, a)$. 
The probability that $P_+ = P[\fr{1}{\sq 2} (|v_{i_k} \ran + | v_{j_k} \ran )]$ clicks is 
\benrr
& & Tr[P_+ \si(G, a)] = \sum^n_{i=1} \lan v_i | P_+ \si(G, a) | v_i \ran \\
& & = \fr{1}{2d_{(G, a)}} \{ \sum^m_{\ba{c} \ell = 1 \\ \ell \ne k \ea} a_{i_\ell j_\ell}[\del_{i_k i_\ell} - \del_{i_k j_\ell} + \del_{j_ki_\ell} - \del_{j_kj_\ell}]^2  + \sum^s_{t=1} a_{i_t i_t}(\del_{i_ti_k} +\del_{i_tj_k})^2\}~~~~~~~~~~~ {(30)} 
\eenrr

The state after the measurement is $P [ \fr{1}{\sq 2} (|v_{i_k} \ran + | v_{j_k} \ran]$.  Let $U^+_{k \ell} $ and $U^+_{kt}$ be $n \times n$ unitary matrices such that $U^+_{k\ell} [ \fr{1}{\sq 2} (| v_{i_k} \ran + | v_{j_k} \ran)] = \fr{1}{\sq 2} (|v_{i_\ell} \ran - | v_{j_\ell} \ran)$ for $\ell = 1, \cdots, k -1, k+1, \cdots, m$ and $U^+_{kt} [ \fr{1}{\sq 2} (| v_{i_k} \ran + | v_{j_k} \ran )] = | v_{i_t} \ran, t = 1, \cdots, s$. Now, with probability $2a_{i_\ell j_\ell}/(d_{(G, a)} - 2a_{i_kj_k})$
we apply $U^+_{k \ell}$ on $P [ \fr{1}{\sq 2} (|v_{i_k} \ran + | v_{j_k} \ran]$ for each $\ell = 1, \cdots, k -1, k+1, \cdots, m$ ,and with probability $a_{i_ti_t}/(d_{(G, a)} - 2a_{i_kj_k})$ we apply $U^+_{k t} $ on $P [ \fr{1}{\sq 2} (|v_{i_k} \ran + | v_{j_k} \ran]$ for each $t = 1, \cdots, s$. Finally we obtain $\si(( G, a) - \{ v_{i_k} v_{j_k} \})$ with probability given by $(30).$ The probability that $ P[  \fr{1}{\sq 2} (|v_{i_k} \ran - | v_{j_k} \ran )]$ clicks is 
\benrr
& &  \fr{1}{2d_{(G, a)}} \{ \sum^m_{\ba{c} \ell = 1 \\ \ell \ne k \ea} a_{i_\ell j_\ell}[\del_{i_k i_\ell} - \del_{i_k j_\ell} - \del_{j_ki_\ell} + \del_{j_kj_\ell}]^2  + \sum^s_{t=1} a_{i_t i_t}(\del_{i_ti_k} - \del_{i_tj_k})^2\} ~~~~~~~ (31)
\eenrr
the state after measurement is $ P[  \fr{1}{\sq 2} (|v_{i_k} \ran - | v_{j_k} \ran )]$. Let $U^-_{k \ell}$ and $U^-_{k t}$ $n \times n$ unitary matrices such that 
$$ U^-_{k \ell} \fr{1}{\sq 2} (|v_{i_k} \ran - | v_{j_k}\ran) = \fr{1}{\sq 2} (|v_{i_\ell} \ran - | v_{j_\ell} \ran)$$
for $\ell = 1, \cdots, k-1, k+1, \cdots, m$ and 
$$ U^-_{k t} \fr{1}{\sq 2} (|v_{i_k} \ran - | v_{j_k}\ran) = |v_{i_t} \ran$$ for $t=1,.....\cdot,s$. With probability $2a_{i_\ell j_\ell}/(d_{(G, a)} - 2a_{i_kj_k})$
 we apply $U^-_{k\ell}$ on $P[\fr{1}{\sq 2} (| v_{i_k} \ran - | v_{j_k} \ran )]$ for each $\ell = 1, \cdots, k-1, k+1, \cdots, m$ and with probability $a_{i_ti_t}/(d_{(G, a)} - 2a_{i_kj_k})$ we apply $U^-_{kt}$ on $P[ \fr{1}{\sq 2} (| v_{i_k} \ran - | v_{j_k} \ran)]$ for each $t = 1, 2, \cdots, s$.  Finally we obtain $\si(( G, a) - \{ v_{i_k} v_{j_k} \})$ with probability given by $(31)$.

The probability that $P[| v_i\ran]$ where $i \ne i_k,j_k$ and $i = 1, \cdots, n$ clicks is 
$$ \fr{1}{d_{(G, a)}}\left\{ \sum^m_{\ba{c} \ell =1 \\ \ell \ne k\ea} a_{i_\ell j_\ell} (\del_{i i_\ell} - \del_{ij_\ell})^2 + \sum^s_{t=1} a_{i_ti_t}(\del_{ii_t})^2 \right\} \eqno{(32)}$$ 
and the state after measurement is $P[|v_i\ran]$. Let $U_{i \ell} $ and $U_{it}$ be $n \times n$ unitary matrices such that $U_{i \ell}[ | v_i\ran] = \fr{1}{\sq 2} (| v_{i_\ell} \ran - | v_{j_\ell}\ran]$ for $\ell = 1, \cdots k-1, k+1, \cdots, m$ and $U_{it} [| v_{i} \ran] = | v_{i_t} \ran$ for $t = 1, \cdots, s$.  With probability $2a_{i_\ell j_\ell} /(d_{(G, a)} - 2a_{i_k j_k})$ we apply $U_{i\ell}$ on $P[| v_i\ran]$ for each $\ell = 1, \cdots, k-1, k+1, \cdots, m$ and with probability $a_{i_t i_t}/(d_{(G, a)} - 2a_{i_k j_k})$ we apply $U_{it}$ on $P[|v_i\ran]$ for each $t = 1, \cdots, s$.

We obtain $\si((G, a) - \{ v_{i_k}, v_{j_k} \})$ with probability given by (32).  This completes the process.

The set of Kraus operators that realizes the TPCP for deleting the edge $\{ v_{i_k}, v_{j_k}\}$ is then 
\benrr
& & \{ \sq{\fr{2a_{i_\ell j_\ell}}{d_{(G, a)} - 2a_{i_kj_k}}} U^+_{k\ell} P[ \fr{1}{\sq 2} (|v_{i_k} \ran + | v_{j_k}\ran)] ;
 \ell = 1, \cdots, k-1, k+1, \cdots, m \} \\
& & \cup \{\sq{\fr{a_{i_ti_t}}{d_{(G,a)} - 2a_{i_kj_k}}} U^+_{kt} P[\fr{1}{\sq 2} (|v_{i_k} \ran + | v_{j_k}\ran)]: t = 1, \cdots, s \} \\
& & \cup \{ \sq{\fr{2a_{i_\ell j_\ell}}{d_{(G, a)} - 2a_{i_kj_k}}} U^-_{k\ell} P[\fr{1}{\sq 2} (|v_{i_k} \ran - | v_{j_k}\ran)]: \ell  = 1, \cdots, k-1, k+1, \cdots m\}\\ 
& & \cup \{ \sq{\fr{a_{i_ti_t}}{d_{(G, a)} - 2a_{i_kj_k}}} U^-_{kt} P[\fr{1}{\sq 2} (|v_{i_k} \ran - | v_{j_k}\ran)] : t = 1, \cdots, s\}\\
& & \cup \{ \sq{\fr{2a_{i_\ell j_\ell}}{d_{(G, a)} - 2a_{i_k j_k}}} U_{i\ell} P[ | u_i\ran] : i = 1, \cdots, n, i \ne i_k, j_k ; \ell = 1, \cdots, k-1, k+1, \cdots, m\} \\
& & \cup \{ \sq{\fr{a_{i_t i_t}}{d_{(G, a)} - 2a_{i_kj_k}}} U_{it} P[|v_i\ran] : i = 1, \cdots, n, i \ne i_k, j_k; t = 1, \cdots, s \} 
\eenrr
The set of Kraus operators that realizes TPCP for adding back edge $\{ v_{i_k}, v_{j_k} \}$ to $(G, a) - \{ v_{i_k} v_{j_k} \}$ is.
\benrr
& & \left\{ \sq{\fr{2a_{i_\ell j_\ell}}{d_{(G, a)} + 2a_{i_kj_k}}} V^+_{k\ell} P[\fr{1}{\sq 2} (|v_{i_k}\ran + |v_{j_k}\ran)] : \ell = 1, 2, \cdots, m\right\} \\
& &\cup \left\{ \sq{\fr{a_{i_ti_t}}{d_{(G, a)} + 2a_{i_kj_k}}} V^+_{kt} P[\fr{1}{\sq 2} (|v_{i_k}\ran + | v_{j_k}\ran)] : t = 1, \cdots, s\right\} \\
& & \cup \left\{ \sq{\fr{2a_{i_\ell j_\ell}}{d_{(G, a)} + 2a_{i_kj_k}}} V^-_{k\ell} P[\fr{1}{\sq 2} (|v_{i_k} \ran - |v_{j_k}\ran)] : \ell = 1, 2, \cdots, m\right\} \\
& & \cup \left\{ \sq{\fr{a_{i_ti_t}}{d_{(G, a)} + 2a_{i_kj_k}}} V^-_{kt} P[ \fr{1}{\sq 2} (|v_{i_k}\ran - |v_{j_k}\ran)] : t = 1,  \cdots, s\right\}\\
& & \cup \left\{ \sq{\fr{2a_{i_\ell j_\ell}}{d_{(G, a)} + 2a_{i_kj_k}}} V_{i\ell} P[|v_i\ra] : i = 1, 2, \cdots, n, i \ne i_k, j_k, \ell = 1, 2, \cdots, m\right\}\\
& & \cup \left\{ \sq{\fr{a_{i_ti_t}}{d_{(G, a)} + 2a_{i_kj_k}}} V_{it} P[|v_i\ran] : i = 1, 2, \cdots, n, i \ne i_k, j_k, t = 1, 2, \cdots, s\right\}
\eenrr
where $V^+_{k\ell}, V^-_{k\ell}, V^-_{kt}, V_{i\ell}, V_{it}$ are $n \times n$ unitary matrix defined as follows: 
\benrr
& & V^+_{k\ell} \fr{1}{\sq 2}(|v_{i_k} \ran + | v_{j_k}\ran) = \fr{1}{\sq 2} (|v_{i_\ell}\ran - |v_{j_\ell}\ran), ~~ \mbox{for}~~ \ell = 1, 2, \cdots, m\\ 
& & V^+_{k t} \fr{1}{\sq 2} (|v_{i_k}\ran + |v_{j_k}\ran) = |v_{i_t}\ran,~~ \mbox{for}~~ t = 1, \cdots, s, \\
& & V^-_{k \ell} \fr{1}{\sq 2} (|v_{i_k} \ran - |v_{j_k}\ran) = \fr{1}{\sq 2} (|v_{i_\ell} \ran - |v_{j_\ell}\ran), ~ \mbox{for}~~ \ell = 1, 2, \cdots, m\\ 
& & V^-_{kt} \fr{1}{\sq 2} (|v_{i_k} \ran - |v_{j_k} \ran) = |v_{i_t}\ran) ,~~ \mbox{for}~~ t = 1, \cdots, s \\
& &V_{i\ell} |v_i\ran = \fr{1}{\sq 2} (|v_{i_\ell}\ran - |v_{j_\ell}\ran),~~ \mbox{for}~~ i = 1, 2, \cdots, n, i \ne i_k, j_k, \ell = 1, \cdots, m\\
& & V_{it} |v_i\ran = |v_{i_t}\ran,~~ \mbox{for}~~ i = 1, \cdots, n, i \ne i_k, j_k, t = 1, \cdots, s.
\eenrr

For deleting a loop $\{v_{i_{t'}}, v_{i_{t'}}\}$ a measurement in the basis $\{ | v_i \ran,  i = 1, \cdots,n \}$ is performed on the system prepared in the state $\si(G, a)$.  Then the probability that $P[| v_i \ran]$ clicks for $i = 1, \cdots n$ is 
$$ \fr{1}{d_{(G, a)}} \left\{ \sum^m_{\ell =1} a_{i_\ell j_\ell}(\del_{ii_\ell} - \del_{ij_\ell} )^2 + \sum^s_{\ba{c} t = 1 \\ t \ne t' \ea} a_{i_ti_t} [\del_{ii_t}]^2 \right\}. \eqno{(33)}$$
The state after the measurement is $P[| v_i \ran]$.  Let $U_{i\ell}$ be $n \times n$ unitary matrices such that $U_{i\ell} [|v_i\ran] = \fr{1}{\sq 2} (| v_{i_\ell} \ran - | v_{j_\ell} \ran).$  For $ i = 1, \cdots, m$ and $U_{it}[| v_i \ran] = | v_{i_t} \ran$, for $ t = 1, \cdots, t' -1, t' + 1, \cdots, s$.  With probability $2a_{i_\ell j_\ell} /(d_{(G, a)} - a_{i_{t'}, i_{t'}})$ we apply $U_{i\ell}$ on $P[|v_i\ran]$ for each $\ell = 1, \cdots, m$ and with probability $a_{i_{t}i_{t}} / (d_{(G, a)} - a_{i_{t'}i_{t'}})$ we apply $U_{it}$ on $P[| v_i \ran]$ for each $t = 1, \cdots, t'-1, t'+1, \cdots s$.  We obtain $\si((G, a) - \{ v_{i_{t'}} v_{i_{t'}}\})$ with probability given by (33).

The set of Kraus operators that realizes the TPCP for deleting the loop $\{ v_{i_{t'}}, v_{i_{t'}}\}$ is 
$$ \left\{ \sq{\fr{2a_{i_\ell j_\ell}}{d_{(G, a)} - a_{i_{t'}, i_{t'}}}} U_{i\ell} P[ |v_i \ran ] ~~ i = 1, \cdots, m ,~~ \ell = 1, \cdots, m\right\}$$ 
$$ \cup \left\{ \sq{\fr{2a_{i_t i_t}}{d_{(G, a)} - a_{i_{t'}, i_{t'}}}} U_{it} P[ |v_i \ran ] ~~ i = 1, \cdots, m ,~~ t = 1, \cdots,  t' -1, t'+1, \cdots s\right\}$$ 
The set of Kraus operators that realizes the TPCP for adding the loop $\{ v_{i_{t'}} v_{i_{t'}}\}$ 
$$ \left\{ \sq{\fr{2a_{i_\ell j_\ell}}{d_{(G, a)} + a_{i_{t'}i_{t'}}}} V_{i\ell} P[|v_i\ran] : i = 1, \cdots, n, \ell = 1, \cdots, m\right\}$$
$$ \cup\left\{ \sq{\fr{a_{i_ti_t}}{d_{(G, a)} + a_{i_{t'}i_{t'}}}} V_{it} P[|v_i\ran] : i = 1, \cdots, n, t = 1, \cdots, s \right\}$$
where $V_{i\ell}, V_{it}$ are $n \times n$ unitary matrices define as follows : 
$$V_{i\ell} |v_i\ran = \fr{1}{\sq 2} (|v_{i_\ell}\ran - |v_{j_\ell}\ran),~~ \mbox{for}~~ \ell = 1, \cdots, m, i = 1, \cdots, n $$ 
$$V_{it} |v_i\ran = |v_{i_t}\ran,~~ \mbox{for}~~ t = 1, \cdots, s, i = 1, \cdots, n.$$

\subsection{Deletion and addition of an edge with real weight, which preserves the positivity of the generalized Laplacian}

Let $(G, a)$ be a graph with real weights on its edges not necessarily positive.   We are basically concerned here with the deletion of  $\{ v_i, v_j\}$ with $a_{v_iv_j} > 0$ and the addition of $\{ v_i,v_j\}$ with $a_{v_iv_j} < 0$, because in other cases the positivity of the Laplacian is preserved. We define the sets 
$$ E^+ = \{ \{v_i, v_j\} \in E(G, a),a_{v_iv_j} > 0 \}, \eqno{(34)}$$ 
$$ E^- = \{ \{v_i, v_j\} \in E(G, a)  , a_{v_iv_j} < 0 \} \eqno{(35)}$$ 
and  $E = E^+ \cup E^-$. 

We define a graph operator $\Xi$ as 
$$ \Xi [E] = E \cup \{ \{v_i,v_i\}, \{v_j,v_j\} : a_{v_iv_i} = a_{v_jv_j} = 2|a_{v_iv_j}|~~\mbox{and}~~ \{v_i, v_j\} \in E^- \} \eqno{(36)}$$ 
Suppose we wish to delete a positive weighted edge $\{ v_{i_k}, v_{j_k}\} \in E^+$ then we define the resulting graph as 
$$ \Xi \cL ((G, a) - \{ v_{i_k}, v_{j_k} \} ) $$ 
where the graph operator $\cL$ is defined in (20b). 

For adding a negative weighted edge between $v_i$ and $v_j, i \ne j$, we act on $E(G, a)$ by the appropriate element of the set of operators $\{ \in_{ij}\}, i, j = 1, \cdots, n, i \ne j$ defined as 
%\benrr
$$\in_{ij} [E]  =  E \cup \{ \{ v_i, v_j\}, \{ v_i, v_i\}, \{ v_j, v_j\} : a_{v_iv_j} < 0, a_{v_iv_i} = 2|a_{v_iv_j}| = a_{v_jv_j}\} \eqno{(37)}$$
%\eenrr
To obtain  the set of the corresponding TPCP operators we decompose the resulting graph, $(G', a')$ given by $\Xi \cL((G, a) - \{ v_{i_k}, v_{j_k}\}) (a_{v_{i_k }v_{j_k}} > 0)$ (eq. (36) ) or by $\in_{ij}((G, a) + \{ v_i, v_j \}) (a_{v_iv_j} < 0)$ (eq. (37)) or by $((G, a) - \{ u_{i_k} v_{j_k} \}) (a_{v_{i_k} v_{j_k}} < 0)$ or by $(G, a) + \{ v_i v_j \}) (a_{v_iv_j} > 0)$ into spanning subgraphs determined by the sets $E^+$ and $E^-$ and treat the spanning subgraph corresponding to $E^-$ replace the weights $a_{u_iv_j}$ of edges $\{v_i, v_j\} \in E^-$ by $- a_{v_iv_j}$, so that both the spanning subgraphs have only positive weights. For getting the Kraus operators we go through the following steps.

(a) First we determine the degree sums for the resulting graphs $(G', a')$ in four cases. 

\bd
\i(i) Deletion of a positive weighted edge $\{ u_{i_k}, v_{j_k} \}$ 
$$ d_{(G',a')} = d_{(G, a)} - 2a_{v_{i_kj_k}} - \sum_i a_{ii} + 2 \sum_{\{u_i,v_j\} \in E^-} |a_{v_iv_j}| \eqno{(38)}$$ 

\i(ii) Addition of a positive weighted edge $\{ v_i, v_j\}$.
$$ d_{(G', a')} = d_{(G, a)} + 2a_{v_iv_j} \eqno{(39)} $$ 

\i(iii)  Deletion of a negative weighted edge $\{ v_{i_k}, v_{j_k}\}$ 
$$ d_{(G', a')} = d_{(G, a)} - 2a_{v_{i_k} v_{j_k}} \eqno{(40)}$$

\i(iv) Addition of a negative weighted edge $\{ v_i, v_j\}$ 
$$ d_{(G', a')} = d_{(G, a)} + 2a_{v_i, v_j} + 4|a_{v_i,v_j}| \eqno{(41)} $$
\ed

(b) We construct the Kraus operators separately for $G^+$ and $G^-$ for deleting the same edge $\{ v_i, v_j\}$ from $G^\pm \sqcup \{ v_i, v_j \}$ or adding the edge $\{ v_i, v_j\}$ to $G^\pm$, using the method given in Section 5.1.  However, the probability of applying various unitary operator $U^\pm_{k\ell}$ and $U^\pm_{k\ell}, U_{i\ell}$ and $U_{it}$ is determinate using $d_{(G', a')}$ as in step (a) above. 

(c) Let $\{ A_i\}$ and $\{ B_i\}$ denote the sets of Krous operators for the graph operations on $G^+$ and $G^-$ as described in (b).  Then 
$$ \si(G', a') = \sum_i A_i \si(G, a) A^\dagger_i - \sum_j B_j \si(G, a) B^\dagger_j  \eqno{(42)}$$ 
and 
$$ \sum_i A_i^\dagger A_i - \sum_j B_j^\dagger B_j = I  \eqno{(43)}$$
which can be justified by construction.

We comment here that it is possible to modify the graph,  after deleting a positive edge or adding a negative edge, which can preserve positivity in different ways, leading to different sets of Kraus operators.  The basic idea is to add new loops.  In our method we try to minimize the addition of loops.  Further, in our method we cannot reverse the graph operation for deleting a positive edge or adding a negative edge.  But this is not a problem since the quantum operations given by super operators are, in general, irreversible.

\subsection {\bf Deleting Vertices}

 In order to delete a vertex $v_i$ from a graph (G,a),
\begin{enumerate}
\item[(i)] Delete edges, including loops, on $v_i$, one by one, by successively
applying the procedure in 5.2.  The resulting graph $(G', a')$ has density matrix with $i$-th row and $i$-th column containing all zeroes.

\item[(ii)] We now perform, on $\si(G', a')$, the projective measurement $M = \{ I_n -  P[|v_i\ran], P[|v_i\ran]\}$.  Since $P[|v_i\ran]$ is the matrix with all elements zero except the $i$-th diagonal element, while $\si(G', a')$ as all zeros in $i$-th row and column,the probability that $P[|v_i\ran]$ clicks $ = Tr(\si P[| v_i\ran]) = 0$.  Thus when $M$ is performed on $\si(G', a')',  I_n - P[|v_i\ran]$ clicks with probability one and the state after measurement is $\si(G', a') - \{ v_i\})$ and is the same as $\si(G', a')$ without $i$-th row and $i$-th column. 
\end{enumerate}
\noi {\bf Adding a vertex :} Let $(G, a)$ be a graph on $n$ vertices $v_1, \cdots, v_n$ and $m$ edges $\{ v_{i_k} v_{j_k} \}, k = 1, \cdots, m, i_k \ne j_k$ and $s$ loops $\{ v_{i_t} v_{i_t}\}, t = 1, \cdots, s;  \le i_k, j_k, i_t \le n$.  Consider the following density operator 
$$ \rho = \left( \fr{1}{2} \sum^2_{i=1} b_{ii} P[|u_i\ran]) \otimes (\si (G, a)\right)$$
where $\{ |u_1\ran, |u_2 \ran\}$ form an orthonormal basis of $\mathbb{C}^2$.  We associate vertices $u_i, i = 1, 2$ to the state $|u_i\ran$.  Consider the graph $H = (\{ u_1, u_2\}, \{ \{ u_1,u_1\}, \{ u_2, u_2\}\})$ with associated weights $b > 0$.  It is easy to check that $\si(H, b) = \fr{1}{2} \sum\limits^2_{i=1} b_{ii} P[|u_i\ran]$.  Also observe that 
$$\rho  =  \si((H, b) \boxdot (G, a))  =  \si(H, b) \otimes \si(G, a).$$

Thus $(H, b) \boxdot (G, a)$ is the graph on $2n$ vertices labeled by $u_1 v_1, \cdots,u_1v_n, u_2v_1, \cdots, u_2v_n$ and with $2m$ edges and $2s$ loops (see Section 4.2) $\{ u_1v_{i_1}, u_1v_{j_1}\} \cdots \{ u_1v_{i_m}, u_1 v_{j_m}\}\\ \{ u_2v_{i_1}, u_2v_{j_1}\} \cdots \{ u_2v_{i_m}, u_2v_{j_m}\}$ and loops $\{ u_1v_{i_t}, u_1 v_{i_t}\}, \{ u_2v_{i_t}, u_2v_{i_t}\}, t = 1, \cdots, s$.  So $(H, b) \boxdot (G, a) = (H_1,a_1) \uplus (H_2, a_2)$ where 
$$ (H_1,a_1) = (\{ u_1v_1 \cdots u_1v_n\}, \{ \{ u_1v_{i_1} ,u_1v_{j_1}\} \cdots \{ u_1v_{i_m}, u_1v_j\} \})$$ 
$$ (H_2, a_2) = ( \{ u_2v_1 \cdots u_2v_n\}, \{ \{ u_2v_{i_1}, u_2v_{j_1}\} \cdots \{u_2v_{i_m}, u_2 v_{j_m}\} \})$$ 
We first delete all edges and loops of $(H, b) \boxdot (G, a)$ which are incident to the vertex $u_2v_1 \in V(H_2, a_2)$ as in Section 5.2.  Now we perform the following projective measurement on $\si((H, b) \boxdot (G, a))$
$$ M = \{ I_{2n} - \sum^n_{i=2} P[|u_2 v_i\ran], \sum^n_{i=2} P[| u_2 v_i\ran] \}$$ 
The probability that $I_{2n} - \sum\limits^n_{i=2} P[v_2 v_i\ran ]$ is one and the state after the measurement is $\si((H_1, a_1) + \{ u_2v_1\})$.

 {\it Example (8)} : Consider the graph as given in the Figure (16) , we want  to delete the edge \{1,2\} with positive weight by means of TPCP.
calculate the Kraus operators for $G^+$ and $G^-$ as in section 5.2 ,where $A_i$ for $i=1,...,\cdots 24$ and $B_i$ for $G^-$ ,$i=1,.....,4$
and substitute in the the equation

$$ \si(G', a') = \sum_i A_i \si(G, a) A^\dagger_i - \sum_j B_j \si(G, a) B^\dagger_j$$ 
where
$$ \si(G, a) = \fr{1}{8} \left[ \ba{cccc} 1 & -1 & -1 & 1 \\ -1 & 3 & -1 & -1 \\ -1 & -1 &3 & -1 \\ 1 & -1 & -1 & 1 \ea \right] .$$
we get
$$ \si(G, a) = \fr{1}{10} \left[ \ba{cccc} 2 & 0 & -1 & 1 \\ 0 & 2 & -1 & -1 \\ -1 & -1 &3 & -1 \\ 1 & -1 & -1 & 3 \ea \right] .$$
and
$$ \sum_{i=1}^{24} A_i^\dagger A_i - \sum_{j=1}^4 B_j^\dagger B_j = I$$

\begin{figure}[!h]
\includegraphics[width=2cm,height=2cm]{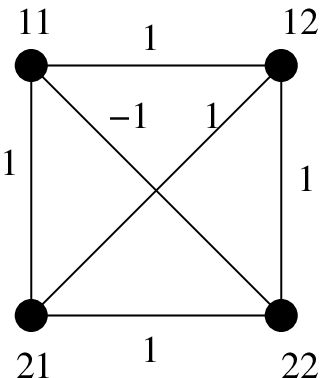}

Figure 16
\end{figure}

\section{Representation of a general hermitian operator by a graph} 

 In this section, we generalize sections 2 - 4, to quantum states in a complex Hilbert space, that is, to the density matrices with complex off-diagonal elements. We have also given rules to associate a graph to a general hermitian operator. We believe that any further advance in the theory reported in this paper will prominently involve graph operators and graphs associated with operators.

\subsection{Representation of a general density matrix with complex off diagonal elements} 

Consider a $n \times n$ density matrix with complex off-diagonal elements.  We associate with this density matrix an oriented  graph $(G, a)$ on $n$ vertices, $m$ edges  and $s$ loops with weight function 
$$ a : V(G) \times V(G) \ra \mathbb{C}.$$

The weight function $a$ has the following properties: 

(i) $a(\{ u, v\}) \ne 0$ if $\{ u, v\} \in E(G, a)$ and $0$ otherwise. 

(ii) $a(\{ u, v \}) = a^*(\{ v, u \})$

we write $a(\{u,v\})\; =\; |a(\{u,v\})| \; e^{i\phi_{uv}}, \phi_{vv} = 0$.

Note that, when $\phi_{ij}=l \pi , l\;=\; 0,1, \cdots$, i.e. $a(\{u,v\})$ is real, positive when $l$ is even and real negative when $l$ is odd.

The degree $d_v$ of vertex $v$ is given by 
$$d_{(G,a)}(v)\;=\;  d_v = \sum_{u \in V(G, a) , \\ u \ne v} |a(\{u,v\})| \;+\; a(\{v,v\}) \eqno{(44)}$$

$$ d_{(G,a)}\;=\; \sum_{v \in V(G, a) } d_v$$

The  adjacency matrix  $M(G, a)$ of a complex  weighted graph with $n$ vertices  is a $n \times n$ matrix whose rows and columns are indexed by vertices in $V(G,a)$ .  
$$M_{uv}\;=\; a(\{u,v\})\;=\;a^*(\{v,u\})\;=\;(M_{vu})^*.$$

The  degree matrix   $\D(G, a)$  of the complex weighted graph is a $n \times n$ real diagonal matrix, whose rows and columns are labeled by vertices in $V(G, a)$ and whose diagonal elements are the degrees of the corresponding vertices. 
$$\D(G, a) = diag [d_v; v \in V(G, a)]$$ 
where $d_v$ is given by equation (44).

The loop matrix $\D_0(G, a)$ of a graph $(G,a)$  is a $n \times n$ real diagonal matrix with diagonal elements equal to the weights of the loops on the corresponding vertices 
$$ [\D_0(G, a)]_{vv} = a_{vv} . $$

The   generalized Laplacian  of a graph $(G, a)$, which includes loops, is 
$$ Q(G, a) = \D(G, a)\;+\; M(G, a) \;-\; \D_0(G, a) \eqno{(45)}$$

Note that $Q(G, a)$  is hermitian matrix . If  the generalized Laplacian $Q(G,a)$ is positive semidefinite, we can define the density matrix of the corresponding graph $(G, a)$ as 
$$ \si(G, a) = \fr{1}{d_{(G, a)}} \; Q(G, a)  \eqno{(46)}$$ 
where $Tr(\si(G, a)) = 1$.

For any $n \times n$ density matrix $\si$ with complex off diagonal elements we can obtain the corresponding graph as follows: 

\noi {\bf Algorithm 6.1 :} 

\bd
\i(i) Label the $n$ vertices of the graph by the kets from the standard orthonormal basis. 

\i(ii) For every nonzero $ij$th element with $j > i$ given by $ a(\{i,j\}) $ draw an edge between vertices labeled $|v_i\ran$ and $|v_j\ran$, with weight $a(\{i,j\})$. 

\i(iii) Ensure that $d_{v_i} = \si_{ii}$ by adding loop of appropriate weight to $v_i$ if necessary. 
\ed
{\it Example (9)} :
(1)
\begin{figure}[!h]
\includegraphics[width=10cm,height=4cm]{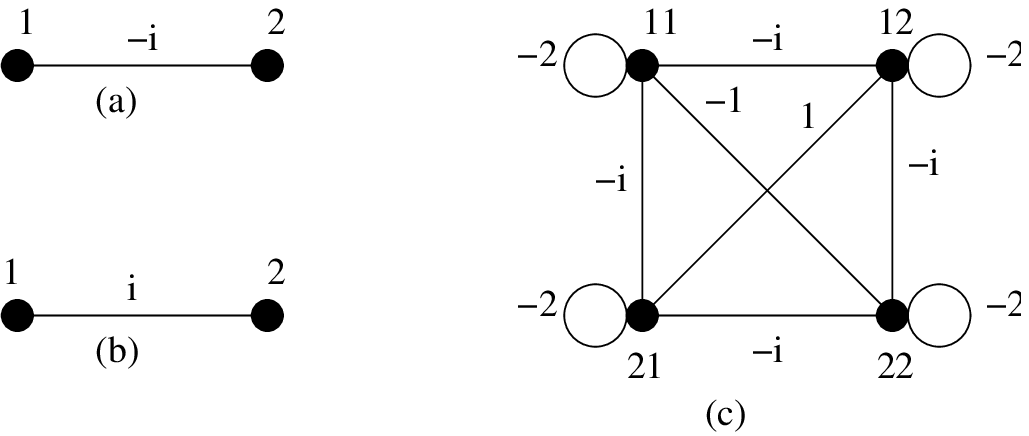}

Figure 17
\end{figure}
$$ P[|y, +\ran] = \fr{1}{2} \left[ \ba{cc} 1 & -i \\ i & 1 \ea \right] = \fr{1}{2} \left[ \ba{cc} 1 & e^{-i\pi/2}\\ e^{-i\pi/2} & 1 \ea \right] $$
where $|y, +\ran = \fr{1}{\sq 2} (|1\ran + i|2\ran)$ and the corresponding graph is as shown in Figure 17a

(2) 
$$ P[|y, - \ran] = \fr{1}{2} \left[ \ba{cc} 1 & i \\ -i & 1 \ea \right] = \fr{1}{2} \left[ \ba{cc} 1 & e^{i\pi/2} \\ e^{-i\pi/2} & 1 \ea \right] $$
where $|y, -\ran = \fr{1}{\sq 2} (|1\ran - i|2\ran)$ and the corresponding graph is as shown in Figure 17b

(3)
$$ P[|y, + \ran|y,+\ran] = \fr{1}{4} \left[ \ba{cccc} 1 & -i & -i & -1 \\ i & 1 & 1 & -i \\i & 1 & 1 & -i\\-1 & i & i & 1\ea \right]  $$
The corresponding graph is as shown in Figure 17c.

Note that Remark 2.1 is valid also for complex weighted graphs.

\textbf{Remark 6.2 :} Theorem 2.3 applies to complex weighted graphs with equation (11) changed to 

$$ \sum^n_{i=1} d^2_i + 2 \sum^m_{k=1} |a_{i_kj_k}|^2 = d^2_{(G, a)} \eqno{(47)}$$ 
  also Lemma 2.4 applies to complex weighted graphs.

\noi {\bf Definition 6.3 :} A graph $(H, b)$ is said to be a factor of graph $(G, a)$ if $V(H, b) = V(G, a)$ and there exists a graph $(H', b')$ such that $V(H', b') = V(G, a)$ and $M(G, a) = M(H, b) + M(H', b')$.  Thus a factor is only a spanning subgraph.  Note that 
$$ a_{v_iv_j} = \left\{ \ba{lll} b_{v_iv_j} & \mbox{if} & \{v_i,v_j\} \in E(H, b) \\ b'_{v_iv_j} & \mbox{if} & \{v_i,v_j\} \in E(H', b') \ea \right. $$ 

Now let  $(G, a)$ be a graph on $n$ vertices $v_1, \cdots, v_n$ having $m$ edges\\ $\{v_{i_1}, v_{j_1}\}, \cdots, \{v_{i_m}, v_{j_m}\}$ and $s$ loops $\{v_{i_1}, v_{i_1}\} \cdots \{v_{i_s}, v_{i_s}\}$ where $1 \le i_1j_1, \cdots, i_m j_m \le n, 1 \le i_1 i_2 \cdots i_s \le n$.

Let $(H_{i_kj_k}, a_{i_kj_k})$ be the factor of $(G, a)$ such that 
$$ [M(H_{i_kj_k}, a_{i_kj_k})]_{u,w} = \left\{ \ba{l} a_{i_kj_k} ~~ \mbox{if}~~ u = i_k~~ \mbox{and}~~ w = j_k ~~\mbox{or}~~  a^*_{i_kj_k} \mbox{if}~~u = j_k, w = i_k \\ 0  ~~ \mbox{otherwise} \ea \right. \eqno{(48)}$$ 
$$ [\D(H_{i_kj_k}, a_{i_kj_k})]_{u,w} = \left\{ \ba{l} |a_{i_kj_k}| ~~ \mbox{if}~~ u = i_k= w ~~\mbox{or} ~~u = j_k= w  \\ 0  ~~ \mbox{otherwise} \ea \right. \eqno{(49)}$$ 

Let $(H_{i_t,i_t}, a_{i_t i_t})$ be a factor of $(G, a)$ such that 
$$ [M(H_{i_ti_t}, a_{i_t i_t})]_{u,w} =  [\D(H_{i_ti_t}, a_{i_ti_t})]_{u,w} = \left\{ \ba{l} a_{i_t i_t}~~ \mbox{when}~~ u = i_t = w \\ 0 ~~ \mbox{otherwise} \ea \right. \eqno{(50)}$$ 

\noi {\bf Theorem 6.4 :} The density matrix of a graph $(G, a)$ as defined above with factors given by equation (48), (49) and (50) can be decomposed as 
$$ \si(G, a) = \fr{1}{d_{(G, a)}} \sum^m_{k=1} 2 |a(\{i_k,j_k\})| \si(H_{i_kj_k}, a_{i_kj_k}) + \fr{1}{d_{(G, a)}} \sum^s_{t=1} a_{i_ti_t} \si(H_{i_ti_t}, a_{i_ti_t}) \eqno{(51)}$$ 
or
$$ \si(G, a) = \fr{1}{d_{(G, a)}} \sum^m_{k=1} 2|a(\{i_k,j_k\})| P[\fr{1}{\sq 2}(|v_{i_k}\ran -e^{i\phi_{i_kj_k}} |v_{j_k}\ran)] + \fr{1}{d_{(G, a)}}  \sum^s_{t=1} a_{i_ti_t} P[|v_{i_t}\ran]\eqno{(52)}$$ 
Where $\phi_{i_kj_k} = \pi$ for any edge  $\{i_k,j_k\}$ with real positive weight and $\phi_{i_kj_k} = 0$  for any real negative weight.

\noi {\bf Proof :} From equation (48), (49), (50) and Remark 6.2, the density matrix 
$$\si (H_{i_kj_k}, a_{i_kj_k}) = \fr{1}{2|a_{i_kj_k}|} [ \D(H_{i_kj_k}, a_{i_kj_k}) + M(H_{i_kj_k},a_{i_kj_k})]$$
is a pure state.  Also, 
$$ \si (H_{i_ti_t}, a_{i_ti_t}) = \fr{1}{a_{i_ti_t}} [ \D_0 (H_{i_t, i_t}, a_{i_ti_t})]$$
is a pure state.  Now 
$$ \D(G, a) = \sum^m_{k=1} \D(H_{i_kj_k}, a_{i_kj_k}) + \sum^s_{t=1} \D_0(H_{i_ti_t}, a_{i_ti_t})$$
$$M(G, a) = \sum^m_{k=1} M(H_{i_kj_k}, a_{i_kj_k}) + \sum^s_{t=1} \D_0(H_{i_ti_t}, a_{i_ti_t}).$$
Therefore , from eq. (46)
%\rr
$$\si(G, a)  =  \fr{1}{d_{(G, a)}} \left[ \sum^m_{k=1} \D(H_{i_kj_k}, a_{i_kj_k}) + \sum^m_{k=1} M(H_{i_kj_k}, a_{i_kj_k})\right]\\
 + \fr{1}{d_{(G, a)}} \left[ \sum^s_{t=1} \D_0 (H_{i_ti_t}, a_{i_ti_t})\right]$$
$$=  \fr{1}{d_{(G, a)}} \sum^m_{k=1} [\D(H_{i_kj_k}, a_{i_kj_k}) + M(H_{i_kj_k}, a_{i_kj_k})] \\
 + \fr{1}{d_{(G, a)}} \sum^s_{t=1} \D_0(H_{i_ti_t}, a_{i_ti_t})$$
$$ =  \fr{1}{d_{(G, a)}} \sum_k 2|a(\{i_k,j_k\})| \si(H_{i_kj_k}, a_{i_kj_k}) \\
 + \fr{1}{d_{(G, a)}} \sum_t a_{i_ti_t} \si(H_{i_ti_t}, a_{i_ti_t}) \eqno{(51)}$$

 In terms of the standard basis, the $uw$-th element of matrices $\si(H_{i_kj_k}, a_{i_kj_k})$ and $\si(H_{i_ti_t}, a_{i_ti_t})$ are given by $\lan v_u | \si(H_{i_kj_k} , ,a_{i_kj_k}) | v_w \ran$ and $\lan v_u | \si (H_{i_ti_t} a_{i_ti_t} | v_w\ran$ respectively.  In this basis 
$$ \si(H_{i_kj_k}, a_{i_kj_k}) = P[ \fr{1}{\sq 2} ( | v_{i_k} \ran -e^{i\phi_{i_kj_k}} | v_{j_k} \ran )]$$ 
$$ \si(H_{i_ti_t}, a_{i_ti_t}) = P[| v_{i_t} \ran ] .$$ 

Therefore equation (51) becomes 
%\benr
$$\si(G, a)  =  \fr{1}{d_{(G, a)}} \sum^m_{k=1} 2|a(\{i_k,j_k\})| P[\fr{1}{\sq 2} (| v_{i_k}\ran -e^{i\phi_{i_kj_k}} | v_{j_k} \ran) + \fr{1}{d_{(G, a)}} \sum ^s_{t=1} a_{i_ti_t} P[ | v_{i_t} \ran]~~~~~~~~~~\eqno{(52)}$$
Where $\phi_{i_kj_k} = \pi$ for any edge  $\{i_k,j_k\}$ with real positive weight and $\phi_{i_kj_k} = 0$  for any real negative weight.

$\hspace{\stretch{1}} \blacksquare$

{\it Example (10)} :

(1) For a graph given in Figure 17b, the density matrix is
\benrr
\si(G, a) & = & \fr{1}{2}\{2 P[\fr{1}{\sq 2} (|1\ran - e^{i\pi/2} |2\ran)]\}=P[\fr{1}{\sq 2} (|1\ran - i|2\ran)]
\eenrr

 (2) For a graph given in Figure 17c,  the density matrix is
\benrr
\si(G, a) & = & \fr{1}{4} \{ 2P[\fr{1}{\sq 2} (|11\ran - e^{-i\pi/2} |12\ran)] + 2P[\fr{1}{\sq 2}(|11\ran - e^{-i\pi/2} |21\ran)]\\
& & + 2 P[\fr{1}{\sq 2} (|11\ran - |22\ran)] + 2 P[\fr{1}{\sq 2} (|12\ran + |21\ran)] \\
& & + 2P[\fr{1}{\sq 2} (|12\ran - e^{-i\pi/2} |22\ran)] + 2P[\fr{1}{\sq 2} (|21\ran - e^{-i\pi/2} |22\ran)] \\
& & - 2P[|11\ran] - 2P[|22\ran] -2P[|12\ran]-2P[|21\ran]\}
\eenrr
$$ \si(G, a) = \fr{1}{4} \left[ \ba{cccc} 1 & -i & -i & -1 \\ i & 1 & 1 & -i \\ i & 1 &1 & -i \\ -1 & i & i & 1 \ea \right] .$$

{\it Example (11)} : Consider the state $$ \si = \fr{1}{3} P[|y, +\ran |y, + \ran] + \fr{2}{3} P[|y, + \ran | \psi \ran]$$ 
where $ |y, +\ran  = \fr{1}{\sq 2} (|1\ran + i|2\ran)$ and $|\psi\ran = \fr{1}{\sq 3} (|1\ran + i\sq 2|2\ran)$ 
\benrr
\si & = & \fr{1}{36} \left[ \ba{cccc} 7 & -(3 + 4\sq 2)i & -7i & -(3 + 4\sq 2)\\ (3 + 4\sq 2)i & 11 & 3 + 4\sq 2 & -11 i \\ 7i & 3 + 4\sq 2 & 7 & -(3 + 4\sq 2)i \\ -(3 + 4\sq 2) & 11i & (3 + 4\sq 2)i & 11 \ea \right] \\
& = & \fr{1}{36} \left[ \ba{cccc} 7 & (3 + 4\sq 2)e^{-i\pi/2} & 7e^{-i\pi/2}  & -(3 + 4\sq 2)\\ (3 + 4\sq 2)e^{i\pi/2}  & 11 & 7 & 11e^{-i\pi/2} \\ 7e^{i\pi/2} & 3 + 4\sq 2 & 7 & (3 + 4\sq 2)e^{-\pi/2} \\ -(3 + 4\sq 2) & 11e^{i\pi/2} & (3 + 4\sq 2)e^{i\pi/2} & 11 \ea \right] 
\eenrr
The corresponding graph is as shown in Figure 18 ,
\begin{figure}[!h]
\includegraphics[width=8cm,height=4cm]{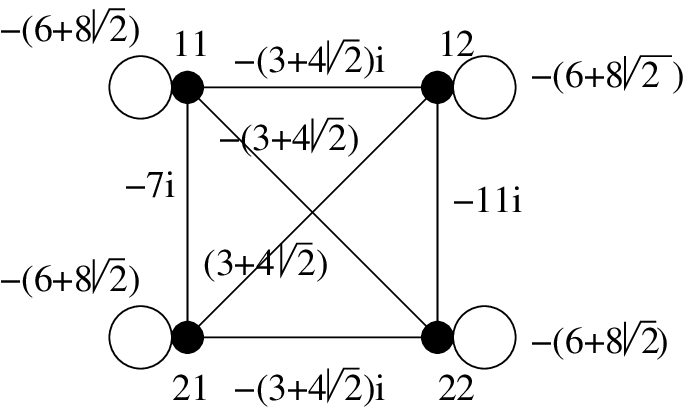}

Figure 18
\end{figure}

and using the equation (45) to get the matrix from graph in Figure 18, 
\benrr
\si(G, a) & = & \fr{1}{36} \{ 2(3 + 4 \sq 2)P[\fr{1}{\sq 2} (|11\ran - e^{-i\pi/2} |12\ran)] \\
& & + 2 \times 7 P[\fr{1}{\sq 2} (|11\ran - e^{-i\pi/2} |21\ran)] + (3 + 4\sq 2) P[\fr{1}{\sq 2}(|11\ran - |22\ran)] \\
& & + 2(3 + 4\sq 2) P[ \fr{1}{\sq 2} (|12\ran + |21 \ran)] + 2 \times 11 P[\fr{1}{\sq 2} (|12\ran - e^{-i\pi/2} |22\ran)] \\
& & + 2(3 + 4\sq 2) P[\fr{1}{\sq 2} (|21\ran - e^{-i\pi/2} |22\ran)] -(6+8 \sq{2})P[|11\ran] - (6 + 8 \sq 2) P[|22\ran]\\
& &  - (6 + 8 \sq 2)P[|12\ran] - (6 + 8 \sq 2) P[|21\ran]\}.
\eenrr
We can check that 
$$ \si(G, a) = \fr{1}{36} \left[\ba{cccc} 7 & -(3 + 4\sq 2)i & - 7i & -(3 + 4\sq 2) \\ (3 + 4\sq 2)i & 11 & 3 + 4 \sq 2 & -11i \\ 7i & 3 + 4 \sq 2 & 7 & -(3 + 4\sq 2)i \\ -(3 + 4\sq 2) & 11i & (3 + 4\sq 2)i & 11 \ea \right]$$ 
We can also check that this state is not pure by applying  Remark 6.2 on the graph.

\subsection{ Separability}

\textbf{Remark 6.6:} The definition of the tensor product $(G,a) \otimes (H,b)$ of two complex weighted graphs $(G,a)$ and $(H,b)$ is the same as given before. However note that $\{v_1,v_2\} \in E(G,a), \{w_1,w_2\} \in E(H,b)$ implies 

$$c(\{(v_1,w_1),(v_2,w_2)\})\;=\; a(\{v_1,v_2\}) b(\{w_1,w_2\})$$ and $$c(\{v_1,w_2),(v_2,w_1)\})\;=\; a(\{v_1,v_2\}) b(\{w_2,w_1\})=a(\{v_1,v_2\}) b^*(\{w_1,w_2\}).$$

\textbf{Remark 6.7 :} Equations (18a) and (18c) are valid for the tensor product of complex weighted graphs. Also, $Q((G,a) \otimes (H,b)) \ne Q(G,a) \otimes Q(H,b)$. Equation (18b) holds good only for graphs without loops, for graphs with only loops or when one factor has no loops and other factor has only loops.
For such graphs  equation (18b) immediately gives $$ d_{(G,a) \otimes (H,b)} (v,w) = d_{(G,a)} (v) \cdot d_{(H,b)} (w).$$
\\
\textbf{6.2.1 Modified tensor product} 
\\
The modified tensor product of two complex weighted graphs requires the operator $\cN$ to be redefined in the following way. We replace the equation (20a) by   
 $$a'_i=\sum_{\begin{subarray}{I}
{v_k \in V(G,a)}\\
\hskip  .4cm {v_k \ne v_i}
\end{subarray}} 
 |a(\{v_i,v_k\})| + a(\{ v_i,v_i\}) \eqno{(53)}$$
The definitions of the operators $\eta, \cL$ and $\Om$ remain the same. Equations (21) to (24) are satisfied by these operators on the complex weighted graphs. We further have 
$$ \ba{ll} \mbox{(i)} & M(\cN \cL(G, a)) = \D(G, a)  - \D_0(G, a) \\ & \D(\cN \cL(G, a)) =  \D(G, a)  - \D_0(G, a) \\ & \D_0(\cN \cL(G, a)) = \D(G, a)  - \D_0(G, a)\\ & Q(\cN \cL(G,a))= \D(G, a)  - \D_0(G, a) \ea \eqno{(54)}$$ 
 The modified tensor product of two complex weighted graphs $(G,a)$ and $(H,b)$ with $p$ and $q(> p)$ vertices respectively is

\benrr
(G,c)=(G, a) \boxdot (H, b) & = & \cL(G, a) \otimes \cL (H, b) \dotplus \cL(G, a) \otimes \cN(H, b) \dotplus  \cN(G, a) \otimes \cL(H, b)\\
& & \dotplus \{ \Om(G, a) \otimes \Om(H, b) \sqcup 2 \cN \cL(G,a) \otimes \cN \cL \eta (H,b)\} ~~~~~~~~~~~~~~{(55)} 
\eenrr 
The weight function $c$ of $(G, a) \boxdot (H, b)$ is obtained via the definition of tensor product and the disjoint edge union.

\textbf{Lemma 6.8 :} $\D((G, a) \boxdot (H, b)) = \D(G, a) \otimes \D(H, b)$.

\textbf{Proof :} Since Lemma 2.10 applies to disjoint edge union of complex weighted graphs,
$$ \D((G, a) \boxdot (H, b) ) = \D( \cL(G, a) \otimes \cL (H, b)) + \D(\cL(G, a) \otimes \cN(H, b)) +\D( \cN(G, a) \otimes \cL(H, b))$$
$$+ \D(  \Om(G, a) \otimes \Om(H, b)) + \D( 2 \cN \cL(G,a) \otimes \cN \cL \eta (H,b))$$
The last two terms are justified because the graphs involved are real weighted graphs. Using Remark 6.7 we get 
 
$$ \D((G, a) \boxdot (H, b) ) = \D( \cL(G, a)) \otimes \D( \cL (H, b)) + \D(\cL(G, a)) \otimes\D( \cN(H, b)) +\D( \cN(G, a)) \otimes \D(\cL(H, b))$$
$$+ \D(  \Om(G, a) ) \otimes \D(\Om(H, b)) + 2 \D( \cN \cL(G,a)) \otimes \D(\cN \cL \eta (H,b))$$

Using equations (21) to (24) and (54), we get, after some simplification, 

 $$\D((G, a) \boxdot (H, b) ) = \D((G, a)) \otimes \D( (H, b))$$
\\
\textbf{Corollary 6.9 :} $d_{(G,a) \boxdot (H,b)}(v,w)\; =\;d_{(G,a)}(v) \cdot d_{(H,b)}(w)$

and$$ d_{(G,a) \boxdot (H,b)}\; =\; d_{(G,a)} \cdot d_{(H,b)}$$

\noi {\bf Proof :}
The first result  follows directly from Lemma 6.8. For the second  note that $$ Tr(\D((G, a) \boxdot (H, b) )) = Tr (\D(G, a) \otimes \D (H, b))= Tr( \D(G, a)) \cdot Tr(\D( H, b))$$ where  $Tr$ denotes the trace.  \hfill $\blacksquare$

\textbf{Theorem 6.10 :} Consider  a bipartite syatem in $\mathbb{C}^p \otimes \mathbb{C}^q$ in the state $ \si $. Then $ \si = \si_1 \otimes \si_2$ if and only if $\si $ is the density matrix of the graph $(G,a) \boxdot (H,b)$ where $(G,a)$ and $(H,b)$ are the graphs having density matrices $\si_1$ and $\si_2 $ respectively.

\textbf{Proof :} \noi {\bf If part :} Given $(G, a), (H, b)$ we want to prove 
$$\si((G, a) \boxdot (H, b)) = \si_1(G, a) \otimes \si_2(H, b).$$

 From the definition of the modified tensor product we can write 
$$ \si((G,a) \boxdot (H, b)) = \fr{1}{d_{(G, a) \boxdot (H, b)}} \{Q[\cL(G, a) \otimes \cL (H, b)$$
$$ \dotplus \cL(G, a) \otimes \cN(H, b)\dotplus  \cN(G, a) \otimes \cL(H, b) \dotplus \{\Om(G, a) \otimes \Om(H, b)) \sqcup 2 \cN \cL(G,a) \otimes \cN \cL \eta (H,b) ]\}$$
Using Remark 6.5 and corollary 6.9   we get \\

$\si((G, a) \boxdot (H, b))  =  \fr{1}{d_{(G, a)} \cdot d_{(H, b)}} [ Q(\cL(G, a) \otimes \cL (H, b))+ Q(\cL(G, a) \otimes \cN(H, b))$
$$+ Q(\cN(G, a)  \otimes \cL(H, b)) + Q(\Om(G, a) \otimes \Om(H, b) \sqcup 2 \cN \cL(G,a) \otimes \cN \cL \eta (H,b))]\eqno{(56)}$$ 

We can calculate every term in (56) using (21) to (24) and (54) and substitute in (56) to get 

$$ \si((G, a) \boxdot (H, b)) = \si(G, a) \otimes \si(H, b) .$$

\textbf{Only if part :} Given $ \si = \si_1 \otimes \si_2$ consider the graphs $(G,a)$ and $(H,b)$ for $\si_1$ and $\si_2$ respectively. Then the graph of $\si$ has the generalized Laplacian $$ Q(G,a) \otimes Q(H,b) = ( \D(G,a)  + M(G,a)- \D_0(G,a)) \otimes (\D(H,b)+M(H,b)- \D_0(H,b))$$
$$  = \D(G,a) \otimes \D(H,b)+ \D(G,a) \otimes (M(H,b)- \D_0(H,b)) +( M(G,a)- \D_0(G,a)) \otimes \D(H,b)+ $$
$$( M(G,a)- \D_0(G,a)) \otimes  (M(H,b)- \D_0(H,b)) \eqno{(57)} $$
Using equation (21) to (24) and (54) we see that RHS of equation (57) is the generalized Laplacian for $(G,a) \boxdot (H,b)$ 
 \hfill $\blacksquare$ 
 
 \textbf{Remark 6.13 :} The proof that the modified tensor product is associative and distributive with respect to the disjoint edge union is the same as that for the case of real weighted graphs (corollary 4.7).
\\
\textbf{Remark 6.14 :} The definition of the cartesian product of graphs is the same as given in definition 4.8.
\\
\textbf{Remark 6.15 :} Corollaries 4.9 and 4.10 apply to complex weighted graphs without any change.

\subsection{Convex combination of density matrices}

Consider two graphs $(G_1, a_1) $ and $(G_2, a_2)$ each on the same $n$ vertices, having $\si(G_1, a_1)$ and $\si(G_2, a_2)$ as their density matrices respectively, where $a_1$ and $a_2$ are complex weight functions. Let $(G,a)$ be the graph of the density matrix $\si(G, a)$ which is a convex combination of $\si(G_1, a_1) $ and $\si(G_2, a_2)$, 
 
$$ \si(G, a) = \la \si(G_1, a_1) + (1 - \la) \si(G_2, a_2), 0 \le \la \le 1.$$

It is straightforward, using the definitions of the operators $\cN ,\cL$ and $\eta$, to varify that

$$ (G,a)=[\la \cN(G_1,a_1)\sqcup (1-\la)\cN(G_2,a_2)]\sqcup [\la \cL(G_1,a_1)\sqcup (1-\la)\cL(G_2,a_2)]$$
$$\sqcup \eta \cL[\la \cL(G_1,a_1)\sqcup (1-\la)\cL(G_1,a_2)] \eqno{(58)}$$
We can apply this equation to any convex combination of density matrices. Let $$ \si(G, a) = \sum_i p_i \si(G_i, a_i) , \sum_i p_i =1$$
Then, 
$$ (G,a)=[\sqcup_i p_i\cN(G_i,a_i)]\sqcup [\sqcup_i p_i \cL(G_i,a_i)] $$
$$\sqcup \eta \cL[\sqcup_i p_i \cL(G_i,a_i)] \eqno{(59)}$$
where $a$ and $\{a_i\}$ are complex weight functions, $a{(\{v_l,v_k\})}=\sum_i a'_i{(\{v_l,v_k\})}$ and $a{(\{v_l,v_l\})}=\sum_i a'_i{(\{v_l,v_l\})}$ with $a'_i=p_i a_i.$

\textbf{Lemma 6.11 :} Let $(G_1,a_1)$, $(G_2,a_2)$ and $(G,a)$ satisfy eq.(58). Then 
$$ \si(G, a) = \fr{d_{(G_1,a_1)}}{d_{(G, a)}}   \si(G_1, a_1) + \fr{d_{(G_2,a_2)}}{d_{(G, a)}}   \si(G_2, a_2).$$ 

\textbf{Proof :} Similar to that of Lemma 2.10.\hfill $\blacksquare$ 

In general, if  $(G,a)$ satisfies eq.(59) for some set of graphs $\{(G_i,a_i)\}$, we have,
$$ \si(G, a) = \fr{1}{d_{(G, a)}} \sum_i d_{(G_i,a_i)}  \si(G_i, a_i).\eqno{(60)}$$
 
\textbf{Theorem 6.12 :} Every graph $(G,a)$ having density matrix $\si(G,a)$ can be decomposed as in eq.(59), where $\si(G_i,a_i)$ is a pure state.

\textbf{Proof :} Same as that of Theorm 2.12.
\hfill $\blacksquare$

\textbf{Corollary 6.16 :} A state of a $k$-partite system is separable if and only if the graph $(G,a)$ for $\si$ has the form $$ (G,a)=[\sqcup_i \cN \boxdot^k_{j=1} (G^j_i, a^j_i)]\sqcup [\sqcup_i  \cL\boxdot^k_{j=1} (G^j_i, a^j_i)] $$
$$\sqcup \eta \cL[\sqcup_i \cL \boxdot^k_{j=1} (G^j_i, a^j_i)] \eqno{(61)}$$

\textbf{Proof :} Same as of Corollary 4.11, where we refer to Theorem 6.4 instead of Theorem 4.5 and Lemma 6.11 instead of Lemma 2.10 and eq.(60) instead of Remark 2.11.\hfill $\blacksquare$ 

Corolary 6.16 says that Werner's definition [1] of a separable state in  $\mathbb{C}^{q_1} \otimes \mathbb{C}^{q_2} \otimes \mathbb{C}^{q_3} \otimes \cdots \otimes \mathbb{C}^{q_k}$ system,  can be expressed using corresponding graphs.

\subsection{Representation of a hermitian operator (observable) by a graph}
In order to represent a general hermitian matrix corresponding to a quantum observable $A$ we lift the requirement that the Laplacian be positive semidefinite and $Tr[A]=1$. In other words we take the generalized Laplacian as the matrix for the graph.

Given a Hermitian matrix $A$, the algorithm 6.1 can be implemented to get its graph $(G, a)$.  The corresponding observable $\h A$ of a graph $(G, a)$ is 
$$ \h A = \sum^m_{k=1} 2a_{i_kj_k}  P[\fr{1}{\sq 2} (|v_{i_k}\ran - e^{i\phi_{i_kj_k}}|v_{j_k}\ran)] + \sum^s_{t=1} a_{i_ti_t} P[|v_{i_t}\ran] \eqno{(58)}$$ 

 {\it Example (12)} :  Give the graph of $\si_x$ and $\si_y$.

(1)~~~~ $\si_x = \left[ \ba{cc} 0 & 1 \\ 1 & 0 \ea \right] $.

The corresponding graph of $\si_x$ is shown in Figure 19a

(2)$ \si_y = \left[ \ba{cc} 0 & -i \\ i & 0 \ea \right] = \left[ \ba{cc} 0 & e^{-i\pi/2} \\ e^{i\pi/2} & 0 \ea \right].$

The corresponding graph of $\si_y$ is shown in Figure 19b

\begin{figure}[!h]
\includegraphics[width=8cm,height=1cm]{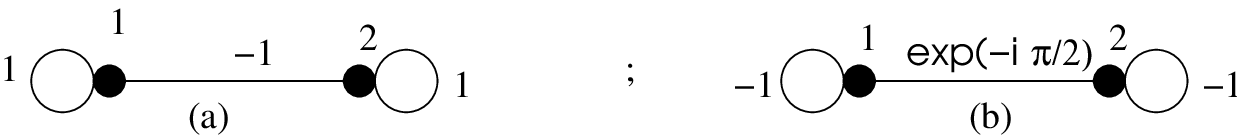}

Figure 19
\end{figure}

Using Equation (46) to get the operators from graphs 

$$ \si_x = -2P[\fr{1}{\sq 2} (|1\ran - |2\ran)] + P[|1 \ran] + P[|2\ran] = |1\ran \lan 2| + |2\ran \lan 1| = \left[ \ba{cc} 0 & 1 \\ 1 & 0 \ea \right] $$ 
$$ \si_y = 2P[\fr{1}{\sq 2} (|1\ran - e^{-i\pi/2}  |2\ran)] - P[|1 \ran] - P[|2\ran] = -i|1\ran \lan 2| + i|2\ran \lan 1| = \left[ \ba{cc} 0 & -i\\ i& 0 \ea \right] $$

\section{Some graphical criteria for the positive semidefiniteness of the associated Laplacian} 
 
 In this section we address the following question. Given a graph, can the positive semidefiniteness of the associated Laplacian be determined using the topological properties of the graph? A general answer to this question seems to be difficult because the theory of weighted graphs, with negative and complex weights is almost unavailable. Many results obtained for simple graphs do not apply to the weighted graphs with real or complex weights.  Nevertheless, we give here the above mentioned criteria in some special cases.

{\bf Lemma 7.1 :} Let $(G,a)$ be a graph with real or complex weights, having one or more nonisolated vertices with degree zero. Then the Laplacian of $(G,a)$ is not positive semidefinite.
 
\noi {\bf Proof :} Such a graph $(G,a)$ has one or more zeroes along the diagonal of its Laplacian with nonzero entries in the corresponding rows. However, a hermitian matrix with one or more zeros in its diagonal has at least one negative eigenvalue unless all the elements in the corresponding rows and columns are zero [18].\hspace{\stretch{1}}$ \blacksquare$

{\bf Lemma 7.2 :} Let $(G,a)$ be a $n$ vertex graph with real weights, having at least one loop and  let the weights on all the loops be negative. Then the Laplacian of $(G,a)$ is not positive semidefinite.

\noi {\bf Proof :} For the given $(G,a)$ and some $x$ in $\mathbb{R}^n$ we have
$$x^T[Q(G,a)]x=\sum_k a_{i_kj_k}(x_{i_k}-x_{j_k})^2-\sum_t|a_{i_ti_t}|x^2_{i_t}$$
where the first sum is over edges and the second sum is over loops. It is easy to check that 
$x^T[Q(G,a)]x<0$ for $x=(1~1~1~\dots  ~1)^T$.\hspace{\stretch{1}}$ \blacksquare$

{\bf Lemma 7.3 :} Let $(G,a)$ be a graph without loops satisfying $a(u,v)=a_{uv}e^{i\phi_{uv}}, (\phi_{uv}\neq 2\pi n)$. Then the associated Laplacian is positive semidefinite.

\noi {\bf Proof :} This follows directly from Theorem 6.4.\hspace{\stretch{1}}$ \blacksquare$

{\bf Observation 7.4 :} Let $(G,a)$ be a graph satisfying $a(u,v)=a_{uv}e^{i\phi_{uv}}$ and $a(\{v,v\})\geq 0$ for all vertices in $V(G,a)$. Then the associated Laplacian is positive semidefinite.

\noi {\bf Proof :} The Laplacian is a hermitian matrix which is diagonally  dominant. Therefore, by Gersgoin circle criterian [14, 15, 19] it is positive semidefinite\hspace{\stretch{1}}$ \blacksquare$

On a $n$ vertex graph $(G,a)$, we define a new graph operator $\Theta(u_i)$ which deletes the vertex $u_i$ and rolls the edges incident on $u_i$ into loops with same weights on the edges connecting neighbors of $u_i$ as shown in Figure 20. We call the resulting subgraph {\it principal subgraph}.  The Laplacian of the principal subgraph obtained by operating $\Theta(u_i)$ on $(G,a)$ is the principal submatrix of the Laplacian of $(G,a)$ obtained by deleting $i$th row and $i$th column.
\begin{figure}[!h]
\begin{center}
\includegraphics[width=8cm,height=3cm]{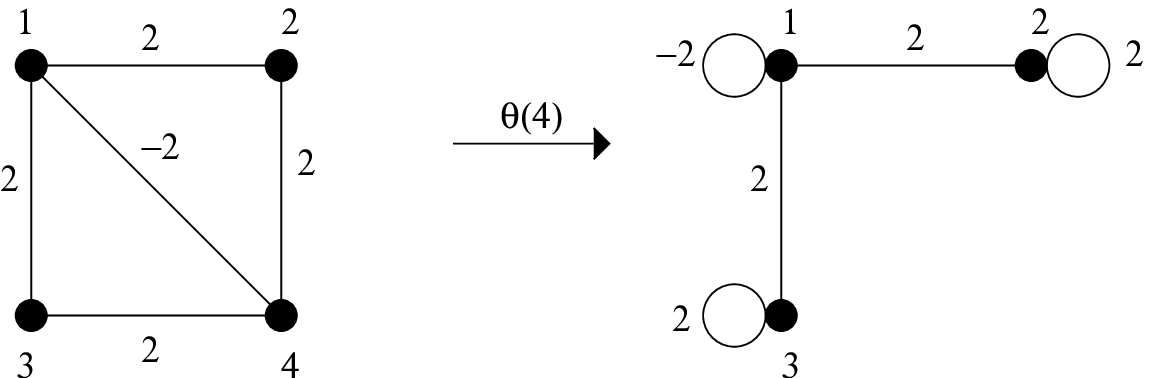}

Figure 20
\end{center}
\end{figure}

{\bf Lemma 7.5 :} If one or more principal subgraphs of $(G,a)$ are not positive semidefinite, then $(G,a)$ is not positive semidefinite.

\noi {\bf Proof :} This follows from the result that all the principal submatrices of a positive semidefinite matrix are positive semidefinite[15].\hspace{\stretch{1}}$ \blacksquare$

{\bf Lemma 7.6 :} Let $(G,a)$ be either a $n$ vertex tree $(n\geq 2)$ or a $n$ vertex cycle $(n\geq 4)$. We assume that there are no loops in $(G,a)$ and that $a(u,v)$ is real for all $\{u,v\} \in E(G,a)$. Then $(G,a)$ has a positive semidefinite Laplacian if and only if $a(\{u,v\})>0$ for all $(u,v)\in E(G,a)$.

\noi {\bf Proof :} \textbf{Only if part :} We prove that $a(\{u,v\})<0$ for any one $\{u,v\} \in E(G,a) \Longrightarrow Q(G,a)\ngeq 0$. Let $(T,a)$ be a tree with $v_1,\dots,v_n$ vertices, and let $\{v_i,v_{i+1}\}$ be an edge in $(T,a)$ with negative weight $a(\{v_i,v_{i+1}\})<0$. We operate on $(T,a)$ by $\Theta(v_{i+1})$. There are two possibilities. If $v_{i+1}$ is a leaf , we get only one component with a negative weighted loop on $v_i$. By Lemma 7.2, the Laplacian of this principal subgraph is not positive semidefinite and by Lemma 7.5 the Laplacian of $(T,a)$ is also not positive semidefinite. If $v_{i+1}$ is not a leaf then $\Theta(v_{i+1})$ will result in two or more principal subgraphs. The principal subgraph containing vertex $v_i$ is a graph having one loop with negative weight. By Lemma 7.2 the Laplacian of this principal subgraph is not positive semidefinite and from Lemma 7.5 $Q(T,a)\ngeq 0.$
Let $C_n$ be an $n$-cycle and let $a(\{v_i,v_{i+1}\})=a<0$. We operate by $\Theta(v_{i+1})$ which results in a $n-1$ vertex path , say $P_{n-1}$ with $v_i$ having a negative loop and $v_{i+2}$ having a positive or negative loop. If both the loops are negative we can use Lemma 7.2 and Lemma 7.5 in succession to show that $Q(C_n,a)\ngeq 0.$
Suppose the loop on $v_{i+2}$ is positive . Then for some $x\in R^{n-1}$ we have
$$x^TQ(P_{n-1},a)x=\sum_{k=1}^{n-2}a_{i_kj_k}(x_{i_k}-x_{j_k})^2+a(v_{i+1},v_{i+2})x_{v_{i+2}}^2-|a|x_{v_i}^2$$
It is straightforward to check that $x^TQ(P_{n-1},a)x<0$ for $x^T=(1~1~\dots 0 ~ 1~\dots ~1)$, that is a vector $x$ with all components $1$ except $v_{i+2}th$ component which is zero. Thus $Q(P_{n-1},a')\ngeq 0$. By Lemma 7.5 $Q(G,a)\ngeq 0$.

\textbf{If part :} Assume $Q(G,a)\ngeq 0$. This implies that there exists at least one $x\in R^n$ satisfying 
$$x^TQ(G,a)x=\sum_k a(i_k,j_k)(x_{i_k}-x_{j_k})^2<0$$

Since $(x_{i_k}-x_{j_k})^2\geq 0$ for all $k$, the above inequality is satisfied only when $a(i_k,j_k)<0$ for some $k$. This proves the if part.\hspace{\stretch{1}}$ \blacksquare$

We observe that the proof of if part applies to all graphs as it should. 

{\bf Lemma 7.7 :} Let all loops on a graph $(G,a)$ have real positive weights. Let every edge $\{u,v\}\in E(G,a)$ having $a(u,v)<0$ be common to pair of $C_3$. Let all such pairs of $C_3$, each containing a negative edge be disjoint. Let all the edges in each pair of $C_3$, other than the contained negative edge have positive weights satisfying $a(u,v)$ greater than the absolute value of the weight on the negative edge. Then the Laplacian of $(G,a)$ is positive semidefinite.

\noi {\bf Proof :} Consider a negative edge common to two $C_3$'s as shown in the Figure 21.

\begin{figure}[!h]
\begin{center}
\includegraphics[width=3cm,height=3cm]{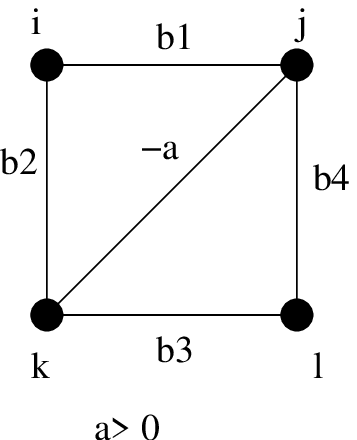}

Figure 21
\end{center}
\end{figure}

By hypothesis $b_i>a,i=1,2,3,4$. We can write $b_i=a+c_i,c_i>0,i=1,2,3,4$. We can decompose this graph as the edge union as shown in Figure 22.
\begin{figure}[!h]
\begin{center}
\includegraphics[width=12cm,height=7cm]{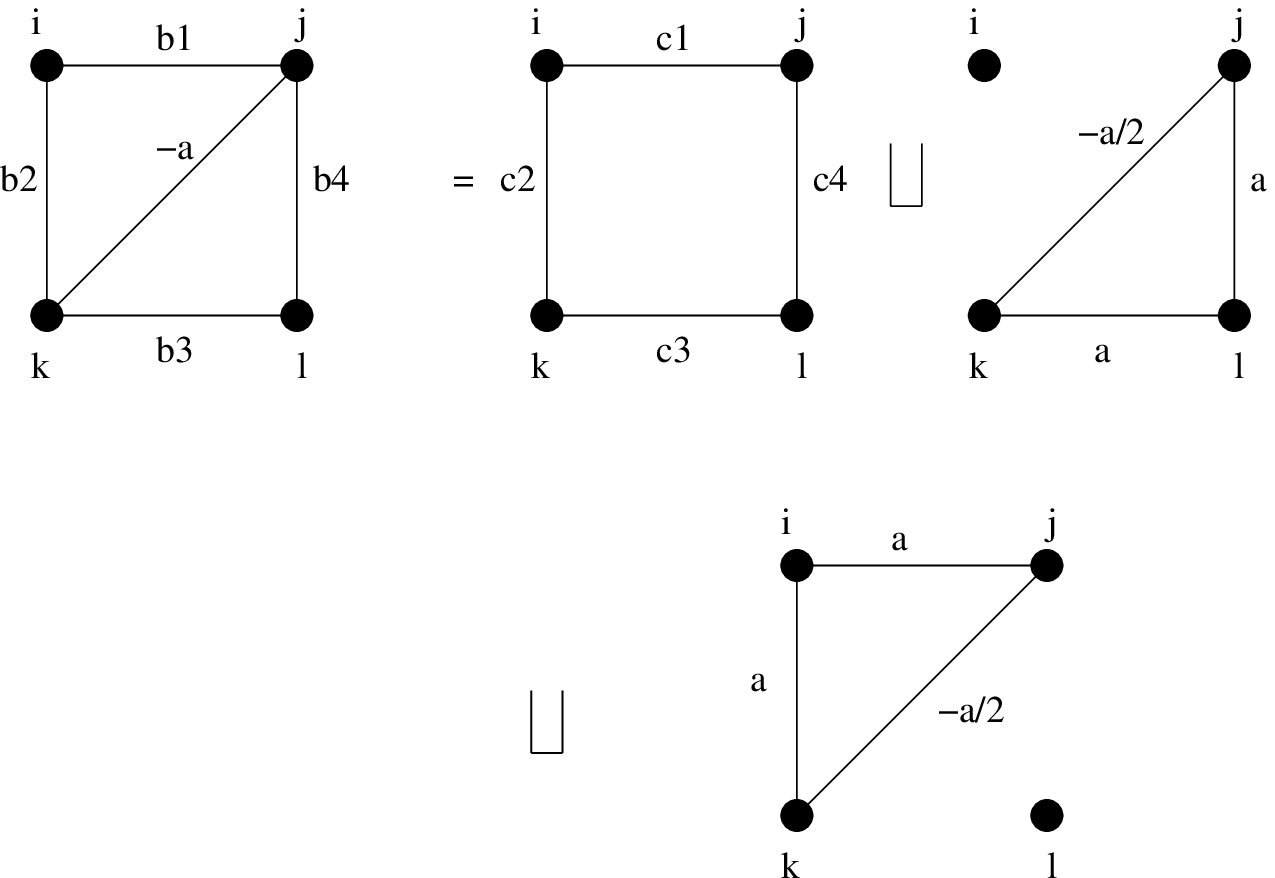}

Figure 22
\end{center}
\end{figure}

The first graph on RHS has all positive weights and hence has a positive semidefinite Laplacian. It is straighforword to check, that the second and the third graphs on RHS correspond to the projectors $P[\fr{1}{\sqrt2}(|j\ran-2|l\ran+|k\ran)]$ and $P[\fr{1}{\sqrt2}(|j\ran-2|i\ran+|k\ran)]$ respectively. Hence they have positive semidefinite Laplacians. The Lapalcian of the graph on LHS is the the sum of the Laplacian of the graphs on RHS (Lemma 2.10), each of which is positive semidefinite. But we Know that the sum of positive semidefinite matrices is a positive semidefinite matrix [15]. Now the graph $(G,a)$ can be written as edge union of the factors (spanning subgraph) as Figure 21 (possibly more than once) and the remaining factor which has all positive weights. The Laplacian of each factor is  positive semidefinite and the Laplacian of the given graph, being the sum of positive semidefinite matrices, is positive semidefinite.\hspace{\stretch{1}}$ \blacksquare$

{\bf Lemma 7.8 :} If all the negative  edges of a real weighted  graph $(G,a)$ occur  as in  the  following   subgraph as shown in Figure (23), where  $c_i>b\; ;\; i= 1, 2, \dots, 8$ and $b>a>0$ then the associated Laplacian is positive semidefinite.
 
\begin{figure}[!h]
\begin{center}
\includegraphics[width=8cm,height=6cm]{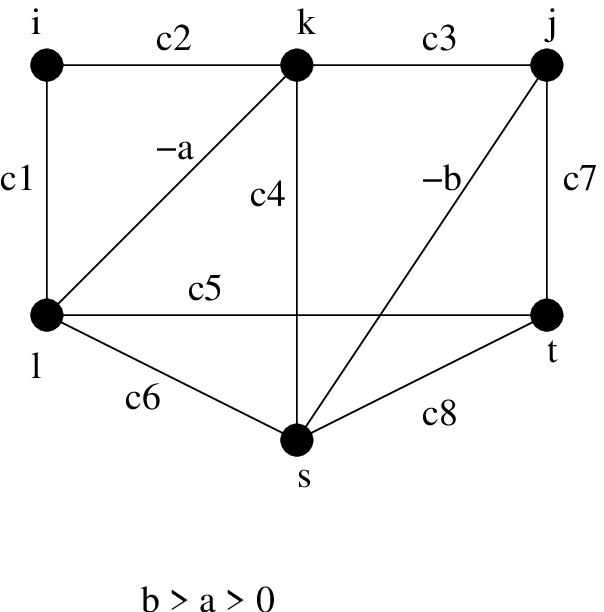}

Figure 23
\end{center}
\end{figure}

\noi {\bf Proof :} We can decompose the above graph into factors as shown in Figure(24)

\begin{figure}[!h]
\begin{center}
\includegraphics[width=14cm,height=8cm]{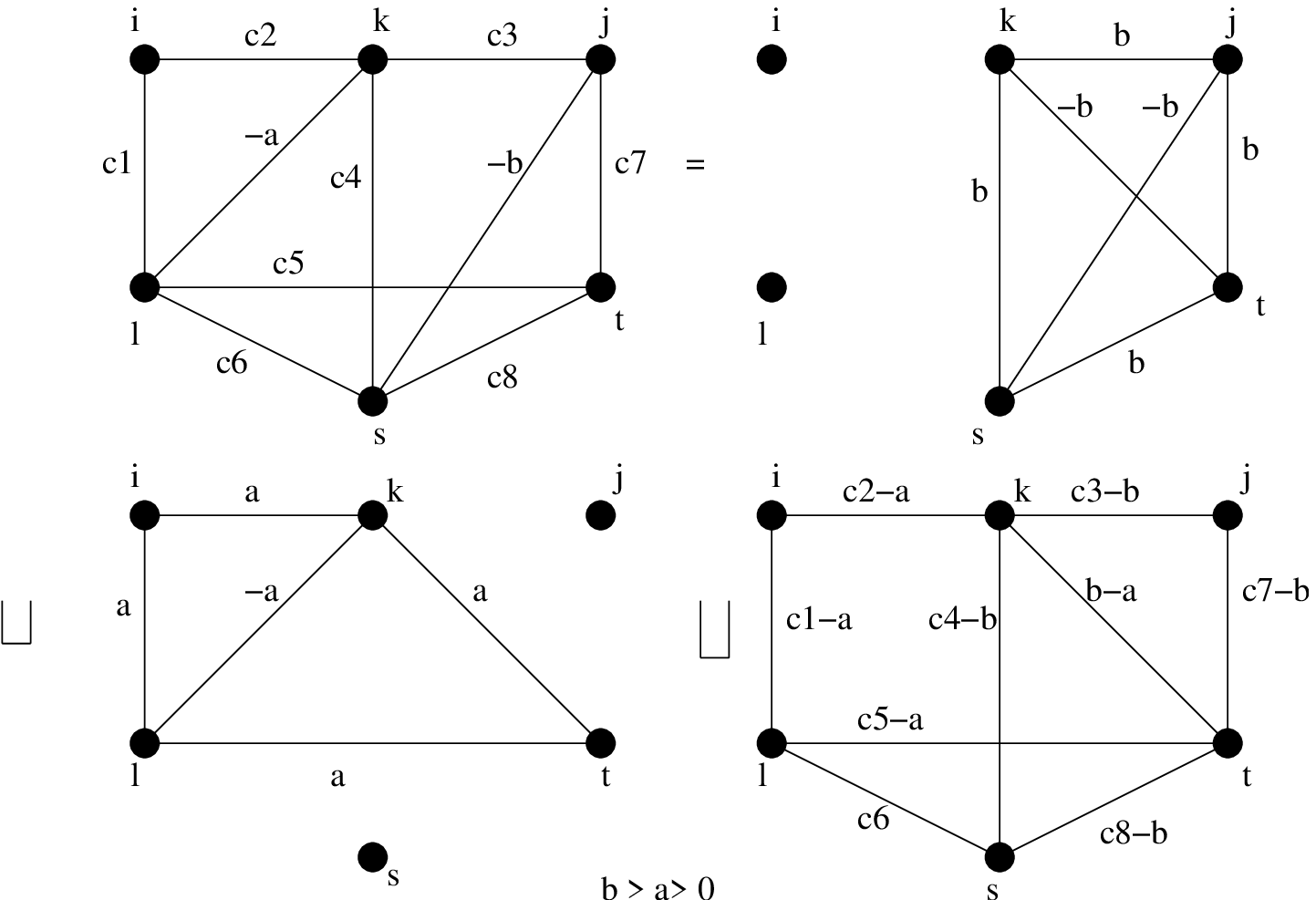}

Figure 24
\end{center}
\end{figure}

From the graphical equation in Figure (24) we see that the first factor on RHS corresponds to $P[|-\ran|-\ran]$, second factor has a positive semidefinite Laplacian from Lemma 7.7 and the third factor has a positive semidefinite Laplacian as it has all positive weights.  Since this graph occurs (once or more) as disjoint subgraphs of $(G,a)$ it can be written as edge union of one or more of these subgraphs and the remaining graph containing only positive or complex edges. Since each of these has a positive semidefinite Laplacian, $(G,a)$ also has a positive semidefinite Laplacian.\hspace{\stretch{1}}$ \blacksquare$

{\bf Lemma 7.9 :} Let $(G^{2^n},a)$ be a complete signed graph with weight function $a_{ij}\in \{-1,1\}$
without loops on $2^n$ vertices $n\geq 1$. Let $E_i$ denote the set of edges incident on $i$th vertex $(|E_i|=2^n-1)$ and let $E_i^+, E_i^-$ denote the sets of edges incident on the the $i$th vertex with weight $+1$ and $-1$ respectively, $(E_i=E_i^+ +E_i^-)$. Let $(G^{2^n},a)$ satisfy the following condition
(i) $|E_i^-|=2^{n-1}-1, i=1, 2, \dots, 2^n$, so that the degree of every vertex$=1$.
Then $(G^{2^n},a)$ corresponds to a pure state in $2^n$ dimensional Hilbert space.

\noi {\bf Proof :} We need to prove that condition (i) in the statement of the Lemma can be realized for all $n$ and that the resulting signed graph corresponds to a pure state for all $n$. We use induction on $n$. It is clear that the assertion is true for $n=1$ with the corresponding pure state given by $P[\fr{1}{\sqrt2}(|1\ran-|2\ran)]$.
Now assume that assertion (that is condition (i) and purity of the corresponding state) is true for $n=k$. For the graph corresponding to $n=k+1$ with $|V(G,a)|=2^{k+1}$ consider the modified tensor product 
\benrr
(G^2,a)\boxdot (G^{2^k},a)=(G^{2^{k+1}},a)=
\{ \cL(G^2, a) \otimes \cL \eta (G^{2^k}, a)\} \dotplus \{\cL(G^2, a) \otimes \cN(G^{2^k}, a)\}\\
   \dotplus \{ \cN(G^2, a) \otimes \cL(G^{2^k}, a)\} \dotplus \{ \Om(G^2, a) \otimes \Om(G^{2^k}, a)\} ~~\mbox{(59)} 
\eenrr
where $\cL, \eta, \cN$ and $\Om$ are graph operators defined in equation (20b) and $G^2,  G^{2^k}$ are graphs with number of vertices  $2$ and $2^k$ respectively.
Note that the last term corresponds to an empty graph as $G^{2^k}$ does not have any loops. Since the modified tensor product of two complete graphs is also a complete graph,  $(G^{2^{k+1}},a)$ is a complete graph. Therefore $|E_i(G^{2^{k+1}},a)|=2^{k+1}-1,  i=1, 2, \dots, 2^{k+1}$.
To show that condition (i) is realized for $(G^{2^{k+1}},a)$ given the induction hypothesis,  we note that the first term in equation (59) contributes $|E^+_i(G^{2^k},a)|$ negative edges to $(1,i)$th vertex in $(G^{2^{k+1}})$ and the third term contributes $|E^-_i(G^{2^k},a)|$ negative edges, while the other two terms have no contribution. Therefore 
$$|E^-_{1i}(G^{2^{k+1}},a)|=|E^+_i(G^{2^k},a)|+|E^-_i(G^{2^k},a)|=2^k-1$$
Similarly, the first three terms contribute $|E^-_i(G^{2^k},a)|, 1$ and $|E^+_i(G^{2^k},a)|$ positive edges to $(1,i)$th vertex. Therefore, 
$$|E^+_{1i}(G^{2^{k+1}},a)|=|E^-_i(G^{2^k},a)|+|E^+_i(G^{2^k},a)|+1=2^{k-1}-1+2^{k-1}+1=2^k$$
and similarly for $E_{2i}.$
That $(G^{2^{k+1}},a)$ corresponds to pure state follows from the fact that the state corresponding to the modified tensor product of two graphs is the tensor product of the states corresponding to the factors. Since the state for $(G^{2^k},a)$ is pure by induction hypothesis and  $(G^2,a)$ is pure, the preceding statement means that $(G^{2^{k+1}},a)$ is a pure product state.\hspace{\stretch{1}}$ \blacksquare$\\

\section{Summary and Comments}
Following is a brief summary of the main features of the paper
\begin{verse} 
(i) We have given rules to associate a graph to a quantum state and a quantum state to a graph, with positive semidefinite generalized Laplacian, for states in real as well as complex Hilbert space(2.1 and 6.1).

(ii) We have shown that projectors involving states in the standard basis are associated with the edges of the graph (Theorems 2.7 and 6.4)

(iii) We have given graphical criteria for a state being pure. In particular we have shown that a pure state must have a graph which is a clique plus isolated vertices  (Theorem 2.3, 2.4, Remark 6.2)

(iv) For states in a real Hilbert space , we have given an algorithm to construct graph corresponding to a convex combination of density matrices, in terms of the graphs of these matrices (2.3.2).

(v) We have defined a modified tensor product of two graphs in terms of the graph operators $\cL, \eta, \cN, \Om$ and obtained the properties of these operators (4.2, 6.2). We have shown that this product is associative and distributive with respect to the disjoint edge union  of graphs (corollary 4.7, Remark 6.11).

(vi) We have proved that the density matrix of the modified tensor product of two graphs is the tensor product matrices of the factors. (Theorem 4.5,  6.10 ). For real density matrices,  we show that a convex combination of the products of density matrices has a graph which is the edge union of the modified tensor products of the graphs for these matrices (corollary 4.11). Thus we can code werner's definition of separability in terms of graphs.

(vii)  We have generalized the separability criterion given by  S. L. Braunstein, S. Ghosh,T. Mansour, S. Severini, R.C. Wilson [2] to the real density matrices having graphs without loops Lemma (4.16).

(viii) We have found the quantum superoperators corresponding to the basic operations on graphs, namely addition and deletion of edges and vertices. it is straightforward to see that all quantum operations on states result in the addition / deletion of edges and/ or vertices , or redistribution of weights. However, addition / deletion of edges / vertices correspond to quantum operations which are irreversible, in general. Hence it seems to be difficult to encode a unitary operator, which has to be reversible, in terms of the operations on graphs. Further , graphs do not offer much advantage for quantum operations which only redistribute the weights, without changing the topology of the graph, as in this case the graph is nothing more than a clumsy way of writing the density matrix.

(ix) In section 6, we generalize the results obtained in sections 2 - 4, to quantum states in a complex Hilbert space, that is, to the density matrices with complex off-diagonal elements. In fact, all the results previous to section 5 go over to the complex case, except Lemma 4.16. Many of these results have been explicitly dealt with (e.g. Theorem 6.4, Remark 6.2, section 6.2 etc). We have also given rules to associate a graph to a general hermitian operator. We believe that any further advance in the theory reported in this paper will prominently involve graph operators and graphs associated with operators.

(x)  Finally, we have given several graphical criteria for the positive semidefiniteness of the generalized Laplacian associated with a graph. Note that by Lemma 7.3 and observation 7.4 all graphs with complex weights, either without loops or  with positive weighted loops have positive semidefinite generalized Laplacians. This characterizes a large class of graphs coding quantum states.

 \end{verse}

 This paper is essentially a generalization of the work by Braunstein, Ghosh and Severini [7] in which the idea of coding quantum mechanics of multipartite quantum systems in terms of graphs was implemented. The motivation in both Braunstein, Ghosh, Severini and this paper is to explore the possibility of facilitating the understanding of mulipartite and mixed state bipartite entanglement using graphs and various operations on them. Whether such a goal can be reached is too early to say. In order to code arbitrary quantum states and observables in terms of graphs, we have to deal with weighted graphs with real or complex weights. Unfortunately, a mathematical theory of such graphs is lacking. Many results pertaining to simple graphs are not available for such weighted graphs. We hope that, through the need of understanding entanglement and related issues the mathematical structure of weighted graphs gets richer and in turn gives a feedback to our understanding of entanglement.

\vspace{0.3cm}

\bc
{\bf Acknowledgement} \\
\ec 

It is a pleasure to acknowledge Sibasish Ghosh for useful discussions and Guruprasad  Kar and Prof. R. Simon for their encouragement.  One of us (ASMH) wishes to acknowledge the Government of Yemen for financial support. We thank Bhalachandra Pujari for his help with Letex and Figures.

\bc
{\bf References} \\
\ec 

\begin{verse} 
[1] Werner R F 1989 {\it Phys.Rev.A}{\bf 40} 4277.\\

[2] Braunstein S L, Ghosh S, Mansour T, Severini S, Wilson R C 2006 {\it Phys. Rev. A}{\bf73} 012320\\

[3] Alber  G, Beth T, Horodecki M, Horodecki P, Horodecki R, Rotteler M, Weinfurter H, Werner R and Zeilinger A 2001  {\it Quantum Information : An Introduction to Basic Theoretical Concepts and Experiments} (Springer Verlag).\\

[4] Bouwmeester D, Ekert A and Zeilinger A 2000 {\it The Physics of Quantum Information} (Springer Verlag).\\

[5] Nielsen A and Chuang I 2000 {\it Quantum Computation and Quantum Information} (Cambridge University Press).\\

[6] Peres A 1993 {\it Quantum Theory : Concepts and Methods} (Kluwer Academic Publishers).\\

[7] Braunstein S, Ghosh S, severini S 2006 {\it Ann. of combinatorics} {\bf 10} No. 3. e-print quant-ph/0406165.\\

[8] Hildebrand R, Mancini S and Severini S arXiv: cs. CC/0607036 (Accepted in Mathematical Structure in Computer Scinence).\\

[9] Kraus K 1983 {\it States, Effects and Operators : Fundamental Notions of Quantum Theory} (Lecture Notes in Physics, Vol. 190, Springer Verlag).\\

[10] Preskill, http://www.theory.caltech.edu/people/preskil/ph2291/. \\

[11] West D 2002 {\it Introduction to graph theory} (Prentice Hall India).\\

[12] Mohar B 1991 {\it The Laplacian spectrum of graphs} (Graph theory combinotorics and Applications vol.II, wiley).\\

[13] Godsil C and Royle G 2001 {\it Algebraic Graph Theory} (Springer-Verlag). \\

[14] Lancaster P and Tismenetsky M 1985 {\it The Theory of Matrices}(Academic Press Inc.).\\

[15] Horn R and Johnson C 1990 {\it Matrix Analysis} (Cambridge University Press).\\

[16] Imrich W and Klavzar S 2000 {\it Product Graphs, Structure and Recognition} (With a forward by Peter Winkler, Wiley - Interscience Series in Discrete Mathematics and Optimization, Wiley - Interscience, New York). \\

[17] Prisner E 1995 {\it Graph Dynamics} (Pitman Research Notes in Mathematics Series, 338, Longman, Harlow).\\

[18] Satake I. 1975 {\it Linear Algebra} (Marcel Dekker, INC. New York).\\

[19] Marcus M, Minc H 1992 {\it A survy of matrix theory and matrix inequalities} (Dover).

\end{verse}

\end{document}